\documentclass[12pt,amssymb,floatfix]{revtex4-1}
\usepackage{graphicx,amsfonts,amssymb,latexsym}
\usepackage{color}

\newcommand{\be}{\begin{equation}}
\newcommand{\ee}{\end{equation}}

\newcommand{\vecp}{{\mathbf p}}
\newcommand{\vecq}{{\mathbf q}}

\newcommand{\x}{{\mathbf x}}

\newcommand{\vl}{{\mathbf l}}

\newcommand{\J}{{\mathbf J}}
\newcommand{\G}{{\mathbf G}}

\newcommand{\W}{{W}}

\newcommand{\mH}{{\mathbf H}}
\newcommand{\mD}{{\mathbf D}}

\newcommand{\mA}{{\mathbf A}}
\newcommand{\mB}{{\mathbf B}}
\newcommand{\mC}{{\mathbf C}}
\newcommand{\mF}{{\mathbf F}}

\newcommand{\mS}{{\mathbf S}}
\newcommand{\mP}{{\mathbf P}}
\newcommand{\M}{{\mathbf M}}

\newcommand{\GGamma}{\mathbf \Gamma}

\newcommand{\Id}{{\mathbf I}}

\newcommand{\der}{\partial}

\newcommand{\GO}{{\mathcal O}}
\renewcommand{\Im}{{\rm Im}}

\newcommand{\vct}[1]{\ensuremath\mbox{\boldmath$ #1 $}}
\newcommand{\Vxi}{{\vct \xi}}
\newcommand{\Veta}{\vct \eta}
\newcommand{\Vzeta}{\vct \zeta}

\newcommand{\matr}[1]{\mathbf #1}
\newcommand{\oper}{\widehat}

\begin{document}

\title{Exact Markovian evolution of quantum systems with several degrees of freedom : Phase space representations}

\author{Aldo R. Fernandes Neto}
\affiliation{Centro Federal de Educa\c{c}\~ao Tecnol\'ogica Celso Suckow da Fonseca, Campus Angra dos Reis, Rua do Areal 522, 23953-030, Angra dos Reis, RJ, Brazil}

\author {Alfredo M. Ozorio de Almeida\footnote{ozorio@cbpf.br}}
\affiliation{Centro Brasileiro de Pesquisas F\'{\i}sicas,
Rua Xavier Sigaud 150, 22290-180, Rio de Janeiro, RJ, Brazil}

\author{Olivier Brodier}
\affiliation{Institut Denis Poisson, Campus de Grandmont, 
Universit\'e de Tours, 37200 Tours, France}

\begin{abstract}

The exact solution of the Lindblad equation with a quadratic Hamiltonian and linear coupling operators
was derived within the chord representation, that is, for the Fourier transform of the Wigner function, also known as the characteristic function.
It is here generalized for several degrees of freedom, so as to provide an explicit expression for the reduced density
operator of any subsystem, as well as moments expressed as derivatives of this evolving chord function.
The Wigner function is then the convolution of its straightforward classical evolution with a widening
multidimensional Gaussian window, eventually ensuring its positivity. Futher on, positivity also holds for the Glauber-Sundarshan
P function, which guarantees separability of the components. In the context of several degrees of freedom, a full dissipation matrix
is defined, whose trace is equal to twice the previously derived dissipation coefficient. This governs the rate at which
the phase space volume of the argument of the Wigner function contracts, while that of the chord function expands.
Examples of Markovian evolution of a triatomic molecule and of an array of harmonic oscillators are discussed.

\end{abstract}

\maketitle

\section{Introduction}

Exact solutions are rare in quantum mechanics and they usually indicate that the system is specially simple.
This is the case of coupled harmonic oscillators, but the situation can become more interesting with an external environment. Indeed, there may be degrees of freedom which are easier to excite experimentally
and are sensitive to the environment, whereas other internal variables are more protected from
decoherence and dissipation. The interaction between these degrees of freedom may then lead to nontrivial behaviour.

More generally, we deal here with the class of symplectic quantum Markovian (SQM) systems that are internally driven
by general quadratic Hamiltonians, whilst coupled implicitly to an external environment
by Lindblad operators \cite{Lindblad,Giulini}, which are linear functions of the position and 
momentum operators. Exact solutions of Lindblad equations within the Wigner-Weyl repesentation 
\cite{Wigner,Gronewold46,Moyal,Report}, or its Fourier transform, the {\it chord representation}, or {\it characteristic function} \cite{Report}, 
were presented in an initial paper \cite{BroAlm04}, henceforth referred to as {\bf I}.
Since the exact quantum solution relies on the classical trajectories of the corresponding Hamiltonian systems
with added dissipation, the solutions for a single degree of freedom can be classified according to the
three generic classes of symplectic dynamical systems for a single degree of freedom,
namely elliptic, hyperbolic and parabolic Hamiltonians. Our objective here is to fully extend
this study to the richer realm of linear systems with several degrees of freedom. We take full advantage of
the exceptional simplicity of the partial trace in the chord representation to obtain directly the evolution of the reduced density operator of any subsystem.

Notwithstanding the strong restriction to SQM systems, there is no constraint on the initial state that 
is then propagated by classical trajectories. In contrast, semiclassical (SC) approximations
that extend SQM evolution beyond the narrow class of quadratic Hamiltonians do rely on special features
of the inital state. A standard option for unitary evolution, constraining the evolution to Gaussian states  
(that is, coherent states or squeezed states) \cite{Hel81} is adapted to Lindblad evolution in \cite{BroAlm10}. 
An alternative treatment by Graefe et al \cite{Graetal18} includes nonlinear Lindblad operators 
within the Gaussian context. However, in general these basis states gradually depart from their Gaussian form,
so that one must further resolve their non-Gaussian evolution again into Gaussian states after finite times. 
%Initial Wigner functions already in a SC form (supported by a lagrangian manifold) \cite{Ber77, AlmHan82}, 
%which are then evolved by SQM systems, have been treated in \cite{OzRiBro, BroOz}.  
%The evolution is straightforward until the classical motion severely distorts the lagrangian manifolds, 
%so as to create caustics. General initial states can also be evolved with the aid of SC propagators, which are again associated to lagrangian manifolds \cite{OzBro06,OzBro11}. 

Isolated symplectic systems with $N>1$ degrees of freedom are usually decomposable into simple generalizations of the three
generic cases of elliptic, hyperbolic or parabolic classical motion. For instance, 
a reaction threshold governed by a saddle point of the Hamiltonian describing a system
with two or more degrees of freedom is resolvable into a combination of elliptic and hyperbolic evolutions.
Williamson's theorem \cite{Arnold} within classical mechanics classifies the possible decompositions 
of a conservative system. It describes the generic normal forms of quadratic classical Hamiltonians 
$H(\x)$, for a physical system with $N$ degrees of freedom, 
defined in the $2N$-dimensional {\it phase space} with coordinates 
$\x = (\vecp,\vecq)=(p_1,...,p_N, q_1,...,q_N)$ as
\be
H(\x) \equiv \frac{1}{2}~\x\cdot \mH~ \x~,
\label{quadraticH}
\ee
where $\mH$ is an arbitrary symmetric matrix. 
Thus, not only is the flow generated by each of these
Hamiltonians a continuous family of {\it symplectic} (i.e. linear canonical) transformations of the phase space, but
there exists an appropriate symplectic transformation that brings the Hamiltonian into its normal form.
%The generic cases contemplated by the normal form for $N>1$ also include possible subspaces with {\it loxodromic Hamiltonians}, 
%describing spiralling hyperbolic evolution for two degrees of freedom, but physically one expects a decomposition
%into the three main types ocurring for $N=1$.
%Williamson's theorem goes on to classify generic families of quadratic Hamiltonians with continuous parameters.

When the quantum Hamiltonian is quadratic, the quantum density operator undergoes a {\it metaplectic transformation} \cite{Voros76,Voros77,Littlejohn86,deGosson06,OAI}. Conveniently, its representation in real phase space, the {\it Wigner function} $W(\x)$ \cite{Wigner,Moyal}, then undergoes a classical symplectic transformation of its variable, generated by the corresponding (\ref{quadraticH}), and so does its Fourier transform, the {\it chord function} $\chi(\Vxi)$ \cite{Report}.
This allows these representations to be transformed exactly to the normal coordinates and back again. %for all the cases classified by Williamson's theorem.

The evolution of the density operator $\hat\rho$ in a Markovian open system 
is determined by the Lindblad equation
\be
\frac{\der\rho}{\der t} = -\frac{i}{\hbar}[\hat{H},\hat{\rho}] 
- \frac{1}{\hbar} \sum_{j=1}^J 2\hat{L}_j\hat{\rho}\hat{L}_j^\dagger -  \hat{L}_j^\dagger  \hat{L}_j\hat{\rho} - 
\hat{\rho} \hat{L}_j^\dagger \hat{L}_j,    
\label{Lindeq1}                                    
\ee
where, together with the Hamiltonian $\hat{H}$, the $J$ Lindblad operators are expressed in the present symplectic case 
as $\hat{L}_j = \vl_j \cdot \hat\x$, in terms of complex $2N$-dimensional vectors $\vl_j=\vl_j' + i \vl_j''$.
Their Wigner-Weyl representation is then just the set of linear functions ${L}_j(\x) = \vl_j \cdot \x$,
which may also be transformed classically by metaplectic similarity transformations. When all the vectors $\vl''_j = 0$, the Lindblad operators are self-adjoint and the dynamics, called "dephasing", has no dissipation. 
What motivates this article is that, although self-adjoint Lindblad operators in (\ref{Lindeq1}) lead to an exact solution which has the same expression as the one described in {\bf I}, the presence of dissipation in the $N>1$ case replaces the scalar {\it dissipative coefficient} 
\be
\gamma \equiv \sum_j  \vl''_j \wedge \vl'_j = \sum_j (\J \vl''_j)\cdot \vl''_j
\label{dissipcoef}
\ee
of {\bf I} by a full dissipative matrix $\GGamma$, leading to a richer behaviour of the solution. Here we used the {\it wedge product} defined by
\be
\Vxi\wedge\x \equiv \sum_n\xi_{pn} q_n - \xi_{qn} p_n = \Vxi_{p}\cdot \vecq - \Vxi_{q} \cdot \vecp \equiv (\J\Vxi) \cdot ~ \x,
\label{wedgedef}
\ee
which also defines the skew matrix $\J$. The trace of $\GGamma$ equals $2\gamma$, so $\gamma$ still can be interpreted as the rate of contraction of the $2N$-dimensional volume of the phase space, in which lies the Wigner function. 

In the following section we extend the exact solution in {\bf I} of the symplectic 
Lindblad equation to systems with several degrees of freedom in the chord representation. This leads in Sec. \ref{momentsW} directly to formulae
for moments, that is, the expectation of products of positions and momenta, and to the propagation of the Wigner function.
It is the convolution of a widening Gaussian window with the Liouville evolution of the original Wigner function, which leads to eventual positivity.
A similar analysis shows in Sec. \ref{positivity} that P positivity, which guarantees
the separability of components, follows after this first threshold is reached. Then, in Sec. \ref{evolution}, 
special features of the chord representation lead to the explicit expressions for reduced density operators.

Since the extention of the results in {\bf I} to several degrees of freedom 
are fairly trivial in the case of mere dephasing, the examples chosen here deal with dissipative environments.
In Sec. \ref{protect} this is a vibrating nonpolar triatomic molecule, which mainly interacts with the environment through its dipolar contribution, that is, its asymmetric vibrational mode. The latter is coupled to the internal symmetric mode by isotopic mass diferences
and its Hamiltonian is derived in Appendix \ref{triat}. In Sec. \ref{network} we treat an array of oscillators coupled with the environment through its surface. Its eigenmodes, calculated in Appendix \ref{eigen}, are shown to be unequally affected by dissipation.

\section{Exact solution of the Lindblad equation with several degrees of freedom}
\label{exact}

Following {\bf I} the exact solution of the Lindblad equation is first derived in the chord representation, 
such that the evolving chord function - or characteristic function
\begin{equation}\label{eq:chord}
\chi(\Vxi,t) = \left[\frac{1}{(2\pi\hbar)^N}\hat \rho(t)\right]_{\chi}(\Vxi) \equiv \int \frac{\mathrm{d}\tilde{\vecq}}{(2\pi\hbar)^N} 
\exp{\left[-\frac{i}{\hbar}\tilde{\vecq}\cdot\Vxi_{p}\right]}
\left< \tilde{\vecq} + \frac{\Vxi_{q}}{2}| \hat{\rho}(t)|  \tilde{\vecq} - \frac{\Vxi_{q}}{2} \right> 
 \, ,
\label{chordef}
\end{equation}
is, by definition, the chord symbol of $\frac{1}{(2\pi\hbar)^N}\hat{\rho}(t)$. Indeed, this is just the symplectic Fourier transform of the evolving Wigner function
\begin{equation}\label{eq:Wigner}
W(\x,t) = \left[\frac{1}{(2\pi\hbar)^N}\hat \rho(t)\right]_W(\x) \equiv \int 
\frac{\mathrm{d}\tilde{\Vxi_q}}{(2\pi\hbar)^N} 
\exp{\left[-\frac{i}{\hbar}\tilde{\Vxi_q}\cdot\vecp\right]}
\left< {\vecq} + \frac{\tilde{\Vxi}_{q}}{2}| \hat{\rho}(t)|  {\vecq} - \frac{\tilde\Vxi_{q}}{2} \right> 
 \, 
\label {Wignerdef}
\end{equation}
which is the Weyl symbol of $\frac{1}{(2\pi\hbar)^N}\hat{\rho}(t)$. The prefactor $\frac{1}{(2\pi\hbar)^N}$ is set to have the integral of $W(\x,t)$ over phase space equal to $1$.
Conversely,
\be
W(\x,t) = \int \frac{\mathrm{d}{\Vxi}}{(2\pi\hbar)^N}~ \exp \left[\frac{i}{\hbar}\x \cdot \J\Vxi \right]\chi(\Vxi,t)~.
\label{FourierWigner}
\ee
The chord function is in general complex, whereas the Wigner function is necessarily real.
For an arbitrary operator $\hat O$ there is no prefactor $\frac{1}{(2\pi\hbar)^N}$ and we will use $O(\x) = \left[\hat O\right]_W(\x)$ and $\tilde{O}(\Vxi)= \left[\hat O\right]_{\chi}(\Vxi)$. 
For typical observables, the Weyl representation equals, at least to first order
in $\hbar$, the corresponding classical phase space function, whereas their chord functions are singular.
With these definitions, the expectation value of an operator is given by phase space integrals in both representations:
\be
\langle \hat{O} \rangle = {\rm tr}~\hat{\rho}~\hat{O} = \int {\rm d}\x ~ W (\x)~O(\x) = \int {\rm d}\Vxi ~ \chi(\Vxi)~\tilde{O}(-\Vxi) .
\label{average}
\ee 
The restrictions to Hermitian operators $\hat{O} = \hat{O}^\dagger$ are $O(\x) = O(\x)^*$ and $\tilde{O}(-\Vxi) = \tilde{O}(\Vxi)^*$,
where $z^*$ denotes complex conjugation of $z$.
	
It is notable that the solution of the Lindblad equation is simplest in its chord representation $\chi$, as long as the Hamiltonian itself remains in its Weyl representation $H$. The unitary part of the Lindblad equation, the Liouville-von Neumann equation, then involves the quadratic function $H(\Vxi)$, instead of $\tilde{H}(\Vxi)$ which is singular,
\begin{eqnarray}
\frac{\der\chi}{\der t}(\Vxi,t)  &=&-\frac{i}{\hbar}[\hat{H},\hat{\rho}]_{\chi}(\Vxi,t) \label{LiouVonNeu} \\ 
 \nonumber
&=& -\frac{i}{\hbar}\int\frac{d\Vxi' d\x'}{(2\pi\hbar)^N}~\exp{\left[\frac{i}{\hbar}\x'\wedge(\Vxi-\Vxi')\right]} ~\chi(\Vxi',t)  
\Big[ H\Big(\x' + \frac{\Vxi}{2}\Big) - H\Big(\x' - \frac{\Vxi}{2}\Big) \Big]    \\ \nonumber
&=& -\J\mH\Vxi \cdot \frac{\der\chi}{\der \Vxi}(\Vxi) = \{H(\Vxi),\chi(\Vxi)\} ~,
\end{eqnarray} 
where the expressions in the last line, bringing in the classical evolution
in terms of the {\it Poisson bracket}, are restricted to the quadratic Hamiltonians (\ref{quadraticH}).
The general integral expression is derived by the mixed product formulae developed in \cite{OzRiBro}.
In the same way, one obtains the Lindblad term $\mathcal L(\hat\rho)=\sum_j \left({2\hat{L_j}\hat{\rho}\hat{L_j}^{\dagger}-\hat{L_j}^{\dagger}\hat{L_j}\hat{\rho}-\hat{\rho}\hat{L_j}^{\dagger}\hat{L_j}}\right)$ as
\begin{eqnarray}
 \left[\frac{1}{(2\pi\hbar)^N}\mathcal L(\hat\rho)\right]_{\chi}(\Vxi,t)
&=&\int \frac{d\Vxi' d\x'}{(2\pi\hbar)^N}~\exp{\left[\frac{i}{\hbar}\x'\wedge(\Vxi-\Vxi')\right]}~\chi(\Vxi',t)   \label{Lind1} \\  \nonumber
 &~& \sum_j \Big[2L_j\left(\x'-\frac{\Vxi}{2}\right)L_j^*\left(\x'+\frac{\Vxi'}{2}\right)\Big. \\ \nonumber
&~& \Big.
-L_j\left(\x'-\frac{\Vxi'}{2}\right)L_j^*\left(\x'-\frac{\Vxi}{2}\right) \Big. \\ \nonumber
&~& \Big. -L_j\left(\x'+\frac{\Vxi}{2}\right)L_j^*\left(\x'+\frac{\Vxi'}{2}\right)\Big].
\end{eqnarray}
In the case of the linear Lindblad operator,
\be 
L_j(\x) = \left[\hat L_j\right]_W(\x) = \vl_j \cdot \x = \vl_j' \cdot \x + i \vl_j'' \cdot \x,
\label{lLin}
\ee
where $\vl_j$ is thus a complex $2N$-dimensional vector, integration leads to
\begin{eqnarray}
 \left[\frac{1}{(2\pi\hbar)^N}\mathcal L(\hat\rho)\right]_{\chi}\left(\Vxi,t\right) 
&=& \frac{1}{2}\sum_j \left[\left(\vl_j'\cdot\Vxi\right)^{2}+\left(\vl_j''\cdot\Vxi\right)^{2}\right]\chi(\Vxi,t) -i\int\frac{d\Vxi' d\x'}{(2\pi\hbar)^N}~ \label{Lind2}
\\  \nonumber
& &   \exp{\left[\frac{i}{\hbar}\J\x'\cdot(\Vxi-\Vxi')\right]}~\chi(\Vxi',t) \sum_j\left[\left(\vl_j''\cdot\Vxi\right)\vl_j'-\left(\vl_j'\cdot\Vxi\right)\vl_j''\right]\cdot\x'   \\ \nonumber
%&=& \frac{1}{2}\sum_j\left[\left(\vl_j'\cdot\Vxi\right)^{2} +\left(\vl_j''\cdot\Vxi\right)^{2}\right]\chi(\Vxi,t) \\ \nonumber
%& & +\hbar~ \sum_j\left[\left(\vl_j''\cdot\Vxi\right)\vl_j'-\left(\vl_j'\cdot\Vxi\right)\vl_j''\right]
%\cdot\mathbf{J}~\frac{\partial\chi}{\partial\Vxi}(\Vxi,t) \\ \nonumber
&=& \frac{1}{2}\left(\Vxi\cdot \mA \Vxi \right)\chi(\Vxi,t)
+ \hbar~(\GGamma \Vxi) \cdot \frac{\partial\chi}{\partial\Vxi}(\Vxi,t) ~,
\end{eqnarray}
defining the {\it decoherence rate matrix} 
\be
\mA = \sum_j \left( \vl_j'~\vl_j'^T + \vl_j''~\vl_j''^T \right), ~~~ \textrm{with} ~~~ \mA^T = \mA
\ee
and the {\it dissipation rate matrix}
\be
\GGamma \equiv \J~ \sum_j \left(\vl_j''~\vl_j'^T-\vl_j'~\vl_j''^T\right), ~~~ \textrm{with} ~~~ \GGamma^T=-\J\GGamma\J.
\label{Gamma}
\ee
When $\J\vl_j'$ and $\J\vl_j''$ are not colinear, $\J\left(\vl_j''~\vl_j'^T-\vl_j'~\vl_j''^T\right)$ is equal to $\gamma_j \Id_2 \oplus {\bf 0}_{2N-2}$, where $\gamma_j = \J\vl_j''\cdot\vl_j'$ and $\gamma_j \Id_2$  acts on the two-dimensional subspace generated by $\J\vl_j'$ and $\J\vl_j''$. Hence $\GGamma$ reduces to $\GGamma = \gamma \Id_2$ in the case of one degree of freedom. 

Finally, the Lindblad equation for a symplectic open system with several degrees of freedom is obtained from (\ref{LiouVonNeu}) and (\ref{Lind2}) as
\begin{eqnarray}
 \frac{\der\chi}{\der t}(\Vxi,t) 
= -\frac{1}{2\hbar}  \left(\Vxi\cdot\mA\Vxi\right) \chi(\Vxi,t) 
-\G_\Gamma \;\Vxi \cdot \frac{\partial\chi}{\partial\Vxi}(\Vxi,t) ~,
\label{Lindeq2}                                
\end{eqnarray}
where the classical evolution matrix
\be
\G_\Gamma = \J\mH + \GGamma, ~~~ \textrm{which~fullfills} ~~~ \G_\Gamma^T = \J \G_{-\Gamma}\J,
\label{defG}
\ee
defines a classical evolution $\Vxi\mapsto e^{t\G_\Gamma}\Vxi$ which is symplectic when $\GGamma=0$.

The solution of (\ref{Lindeq2}) is
\be
\chi(\Vxi,t) = \chi(e^{-t\G_\Gamma}\Vxi,0)~ \exp{\left[-\frac{1}{2\hbar}\Vxi\cdot \M(t)~\Vxi\right]},
\label{solutionj}
\ee
where the positive {\it decoherence matrix} is defined as
\be
\M(t) \equiv \int_0^t dt'~ \left[e^{(t'-t)\G_\Gamma}\right]^T \mA \;e^{(t'-t)\G_\Gamma}.
\label{defM}
\ee
Hence the evolved chord function can be considered as a Gaussian channel \cite{Eis05}.

The nonisotropic dissipation portrayed by $\GGamma$ is the essential new feature of the Markovian
evolution derived in the present work. One verifies that
\be
 \det e^{t\G_\Gamma} =\exp{\rm tr}~[t \G_\Gamma]  
=\exp~\left[\sum_j(\J \vl_j'' \cdot \vl_j' - \J \vl_j' \cdot \vl_j'')t\right] = e^{2 \gamma t},
\label{tracediss}
\ee
since ${\rm tr}~\J\mH = 0$. 
Equation (\ref{tracediss}) shows that a positive {\it dissipation coefficient} $\gamma$ (\ref{dissipcoef}) 
induces an {\it expansion} of the total volume chord space. However a single mode will not necessarily expand, as the real parts of the complex eigenvalues $\lambda_j$ of $\G_\Gamma$ are not necessarily positive.

Let us define $\mP$ such that $\G_\Gamma = \mP \;\mD_{\{\lambda_j\}}\;\mP^{-1}$, where $\mD_{\{\lambda_j\}}$ defines a diagonal matrix with eigenvalues $\lambda_j$. Then
\be
\M(t) = \left(\mP^{-1}\right)^T\; \left( \mB\circ \mC(t)\right) \;\mP^{-1},
\label{explicitM}
\ee
where $\mB\circ \mC(t)$ is the Hadamard product of $\mB$ and $\mC(t)$, that is $\left(\mB\circ \mC(t)\right)_{j,k} = \mB_{j,k}\mC_{j,k}(t)$ with
\begin{eqnarray}
\mB &= \mP^T \mA \mP \label{B}\\
\mC_{j,k}(t) &= \frac{1-e^{-t(\lambda_j+\lambda_k)}}{\lambda_j+\lambda_k}\;,\label{Ct}
\end{eqnarray}
so the time dependence of $\M(t)$ is completely determined by the eigenvalues of $\G_\Gamma$ which appear in the matrix $\mC(t)$. If these eigenvalues have a strictly positive real part, then asymptotically backward evolution drives all chords to the origin, that is 
\be
\mathop{lim}_{t\rightarrow\infty}\chi(e^{-t\G_\Gamma}\Vxi,0) = (2\pi\hbar)^{-N} ~,
\ee
whereas $\M(t)$ converges exponentially to a positive definite constant matrix,
so that
\be
\chi(\Vxi,\infty) = (2\pi\hbar)^{-N} ~ \exp{\left[-\frac{1}{2\hbar}\left(\mP^{-1}~\Vxi\right) \cdot \mB\circ \mC(\infty)\; \left(\mP^{-1}~\Vxi\right)\right]}  
\label{solutionfty}
\ee
 is a multidimensional Gaussian with
\be
\mC_{j,k}(\infty) = \frac{1}{\lambda_j+\lambda_k}.
\label{Cinfini}
\ee
 In the general case, $\mC(\infty)$ is singular when the real part of some $\lambda_j+\lambda_k$ is negative. In all cases, even singular, $M(\infty)$ is positive by construction.

The generic eigenvalues of $\J\mH$ can be pairs $\pm\omega$ or $\pm i\omega$, but also quadruples $\pm g\pm i\omega$, or zero. However, according to Arnold in Appendix 6 of \cite{Arnold}, we can remain sufficiently generic if we restrict to the pairs $\pm\omega$ or $\pm i\omega$, with $\omega\geq 0$, which then includes pairs of zeros. By doing this, we leave aside less generic families of spiraling motion. Let us then see what would give our formalism in this situation, and suppose that $\mH$ has the form, $\mH=\mD_{\omega_1,\omega_1',\omega_2,\omega_2',\ldots,\omega_N,\omega_N'}$, with $\omega_j \geq 0$, and $\omega_j' = \omega_j$ for elliptic modes whereas $\omega_j' = -\omega_j$ for hyperbolic ones. Then, we adopt the typical scenario where each mode $j$ is coupled to environment through one or several $\vl_k$, corresponding to annihilation and creation operators. Under these hypotheses, $\G_\Gamma$ and $\mA$ are both block diagonal with 2x2 blocks, and $\chi(\Vxi,t)$ is simply a product of one degree of freedom chord functions. The blocks of $\G_\Gamma$ have the form
\be
\G_j = \left(\begin{array}{cc}\gamma_j & -\omega_j' \\ \omega_j & \gamma_j \end{array}\right),
\label{blockG}
\ee
and its eigenvalues are $\gamma_j \pm i \omega_j$ for the elliptic modes, $\gamma_j \pm \omega_j$ for the hyperbolic ones, and a pair of $\gamma_j$ in the degenerate case. Notice that if $\omega_j >\gamma_j$ then the direction corresponding to $\gamma_j - \omega_j$ will shrink in the chord space even if the system is dissipative, with other directions expanding. 

The detailed balance condition for the Lindblad equation applies when $\mathcal L(e^{-\beta \widehat H})=0$, which, in the Weyl representation, corresponds to the condition
\be
\GGamma + 2\hbar \J\mA\J\mF = 0,
\ee
where $\mF$ is the matrix such that, as shown in \cite{Hor95},
\be
\left[e^{-\beta \widehat H}\right]_W = e^{-\frac{1}{2}\x\cdot\mF\x}.
\ee
Sticking to an elliptic Hamiltonian, so that the Gibbs state may have a physical meaning, $\mF$ is then diagonal in the Williamson base, with eigenvalues $\frac{\tan{\left(\frac{\beta\hbar\omega_j}{2}\right)}}{\hbar\omega_j}$ instead of $\omega_j$. If each mode couples to the environment through two Lindblads $\widehat a$ and $\widehat a^\dagger$ acting in its subspace, then $\vl_{j,1} = (i u_j,u_j)$ and $\vl_{j,2} = (-i v_j,v_j)$, and the detailed balance condition reads
\be
v_j = \sqrt{\frac{1-2\tan{\left(\frac{\beta\hbar\omega_j}{2}\right)}}{1+2\tan{\left(\frac{\beta\hbar\omega_j}{2}\right)}}}\; u_j = \sqrt{\frac{\bar n_j}{\bar n_j +1}}\; u_j
\ee
where $\bar n_j$ can be interpreted as the average number of photons at equilibrium in mode $j$.

In the general case, unlike the above block diagonal case, the eigenvalues $\lambda_j$ of $\G_\Gamma$ can have a negative real part even in a dissipative case with only elliptic modes. Take for instance $\oper H = \frac{\oper p_1^2}{2} + \oper q_1^2 + \frac{\oper p_2^2}{2} + \frac{\oper q_2^2}{2}$ with $l_1=(i,1,0,\frac{3}{2})$ and $l_2=(0,\frac{3}{2},i,1)$, then $\G_\Gamma$ has eigenvalues $-0.4944\pm 1.2179 i$ and $2.4944\pm 1.2179 i$ while $\gamma_1=\gamma_2=1$.

\section{Moments and the Wigner function}
\label{momentsW}
Before presenting the evolution of the more familiar Wigner function, it should be recalled that
the chord function, or characteristic function, is already a complete representation of the density operator with its own advantages.
The fact that the chord representation of the identity operator $\hat {\Id}$ is a Delta function,
\be
\Id(\Vxi)= (2\pi\hbar)^N\delta(\Vxi),
\label{Id}
\ee
leads to the evolving expectation of (appropriately symmetrized) polynomials of position 
and momentum operators being represented exactly as corresponding polynomials of derivatives of the chord function.
Indeed, one easily obtains from (\ref{average}) the statistical moments 
for a single degree of freedom as
\begin{equation}
\begin{array}{cc}
\langle  q^n \rangle_t & =  \rm tr\> \widehat{q}^n \>\widehat{\rho} = ~
(i~\hbar )^n \ \frac{\partial^n}{\partial \xi_p^n} (2\pi\hbar)^L
\ \chi(\Vxi,t)~ \Big|_{\Vxi = 0} \; \\
\langle  p^n \rangle_t & =  \rm tr\> \widehat{p}^n \>\widehat{\rho} = 
(-i\hbar )^n \ \frac{\partial^n}{\partial \xi_q^n} (2\pi\hbar)^L
\ \chi(\Vxi,t) \Big|_{\Vxi = 0} \ ,
\end{array}
\label{moments}
\end{equation}
just as a classical characteristic function supplies the moments of its parent probability distribution. 
Shifting the phase space origin to 
$\langle \x\rangle$,
we can define $\bf K$ the {\it Schr\"odinger covariance matrix} \cite{Schro30}
just as its classical counterpart, with
$\delta p^2=\langle \widehat{p}^2 \rangle$, $\delta {q}^2=\langle \widehat q^2 \rangle$
and $(\delta pq)^2=\langle (\widehat p \widehat q + \widehat q \widehat p)/ 2 \rangle$ in the case of a single degree of freedom.
It is then obvious that the expansion of the real part of the chord function at the origin is just
\be
{\rm Re}~ \chi(\Vxi)=(2\pi\hbar)^{-N} - \Vxi \cdot {\bf K} \; \Vxi + ...
\ee
and we can interpret the {\it uncertainty},
\be 
\Delta_{\bf K}= \sqrt{\det {\bf K}}, 
\ee
as proportional to the volume of the ellipsoid: $\Vxi\ \cdot {\bf K} \; \Vxi=1$.
Evidently, this volume is invariant with respect to symplectic transformations,
so that $\Delta_{\bf K}$ is a symplectically invariant measure of the uncertainty
of the state.

In \cite{Barthel22}, Barthel and Zhang compute the dynamics of $\bf K$ itself through second quantization, in Markovian open systems with quadratic Hamiltonian and linear Lindblads. Then they have access to the second moments in the bosonic case and in the fermionic case. In comparison, our approach deals only with the bosonic case, but it gives access to higher moments as well. In other words, both approaches are equivalent for Gaussian states in the bosonic case, while \cite{Barthel22} brings in addition a complete description of Gaussian states in fermionic systems, whereas our apporach brings in addition a complete description of non Gaussian states in bosonic systems.

The evolution of the Wigner function follows by inserting (\ref{solutionj}) 
into the symplectic Fourier transform (\ref{FourierWigner}), which gives, defining the original chord
$\Vzeta = e^{-t \G_\Gamma}\Vxi$ and using (\ref{tracediss}),
\be
 W(\x,t) = e^{2\gamma t} \int \frac{d\Vzeta}{(2\pi\hbar)^N}~ \chi(\Vzeta, 0)~
\exp\left[-\frac{1}{2\hbar} \Vzeta \cdot \M(-t)\Vzeta\right] ~\exp\left[-\frac{i}{\hbar}\J\x \cdot e^{t \G_\Gamma}\Vzeta\right] ~.
\label{evolwigeta}
\ee
Then, using relations (\ref{defG}), $\left(e^{t\G_\Gamma}\right)^T = \J e^{t\G_{-\Gamma}}\J$ and $\det \M(-t) = e^{4\gamma t} \det\M(t)$, the expression (\ref{evolwigeta}) can be given the form of the convolution
\begin{eqnarray}
W(\x,t) &=& \frac{1}{\sqrt{\det\M(t)}} \int \frac{d\x'}{(2\pi\hbar)^N}~ \W(\x', 0)  \label{Wigev} \\ \nonumber
& & \exp\left[-\frac{1}{2\hbar} \left(\x'-e^{-t \G_{-\Gamma}}\x\right) \cdot \M_{\J}(-t)^{-1}\left(\x'-e^{-t \G_{-\Gamma}}\x\right)\right] ,
\end{eqnarray}
where $\M_{\J}(t) = -\J \M(t)\J$.

The asymptotic equilibrium Wigner function is directly obtained as the Fourier transform of (\ref{solutionfty}):
\be
W(\x,\infty) = \frac{(2\pi\hbar)^{-N}}{\sqrt{\det \M(\infty)}} ~ \exp \left(-\frac{1}{2\hbar}\x\cdot \M_{\J}(\infty)^{-1}~\x\right) ~.
\ee
This is a positive Gaussian function and so the question is when does it lose its negative regions, which are generally present in the initial pure state.

If the initial state is a coherent state
or its symplectic deformation $|\Veta, {\mathbf S} \rangle$, then the initial Wigner function is the Gaussian
\be
 W_{\Veta,  {\mathbf S}}(\x) = \frac{1}{(\pi\hbar)^N} \exp{\left[-\frac{1}{\hbar}\left({\mathbf S}(\x-\Veta)\right)^2 \right]}
=\frac{1}{(\pi\hbar)^N}\exp{\left(-\frac{1}{\hbar}(\x-\Veta)\cdot{\mathbf S}^T {\mathbf S}(\x-\Veta)\right)}~,
\label{Wcoherent}
\ee
where $\Veta$ is the expectation of $\hat{\x}$ and $\mathbf S$ is a symplectic matrix. Therefore no zero is introduced in the evolved Wigner function (\ref{Wigev}), since  
the convolution of a Gaussian with a Gaussian window is a wider Gaussian.

Being normalized to $1$, a general pure state Wigner function is dominated by positive regions, but it can have oscillating parts with negative regions.
Its convolution (\ref{Wigev}) with a Gaussian that broadens in time then acts to smoothen these oscillations, while gradually ironing out the negative regions. The positivity of the evolved Wigner function 
throughout phase space is then equivalent to the positivity of
\begin{eqnarray}
 W(e^{t\G_{-\Gamma}}\x,t) &=& \frac{1}{\sqrt{\det\M(t)}} \int \frac{d\x'}{(2\pi\hbar)^N} ~ \W(\x', 0)  \label{Wigev2} \\   \nonumber
& & \exp\left[-\frac{1}{2\hbar} (\x'-\x) \cdot \M_{\J}(-t)^{-1}(\x'-\x)\right]\; ,
\end{eqnarray}
which can be compared to the Husimi function, also known as the $Q$-function \cite{Hus40} \cite{Char02},
\begin{eqnarray}
Q(\Veta,0) &=& |\langle\Veta|\psi(0) \rangle|^2 =  \int d\x' ~W(\x',0) ~W_{\Veta,\Id}(\x') 
\label{Husimi} \\ \nonumber
       &=& \int \frac{d\x'}{(\pi\hbar)^N}~ \W(\x', 0) ~\exp\left[-\frac{1}{\hbar} (\x'-\Veta)^2\right] ,
\end{eqnarray}
where $|\psi(0) \rangle$ is the Hilbert space vector corresponding to $W(\x,0)$.
To simplify this comparison, Eq. (\ref{Wigev2}) can be brought close to the Husimi function
by expressing, with the aid of Williamson's theorem \cite{Arnold}, the positive matrix $\M(-t)$ in its normal form, which is a direct sum of $N$ independent harmonic oscillators with frequencies $\Omega_n(t)$. Thus, for each time there exists a symplectic transformation $\x'\mapsto \x'(t)={\mathbf S}(t)^{-1}\x'$ such that
\be
\x'(t)\cdot {\mathbf S}(t)^T\M_{\J}(-t)^{-1}{\mathbf S}(t)\x'(t) = \sum_{n=1}^N \frac{1}{\Omega_n(t)} \left[{p'}_j(t)^2 + {q'}_j(t)^2\right]. 
\ee
In general the pairs of eigenvalues $\Omega_n(t)$ are not equal, so that, even in this priviledged evolving symplectic frame, 
each different mode will generally be identified with a Husimi function at a different instant. 
Nonetheless, one can establish bounds for overall positivity, by using the same argument as in Sec. IV of {\bf I}. 

We define $\Omega_{\textrm{min}}(t)$ as the smallest eigenvalue and $\Omega_{\textrm{max}}(t)$ as the largest one. The eigenvalues are increasing functions of time, so let us suppose that there exists $t_-$ such that $\Omega_{\textrm{max}}(t) \leq \frac{1}{2}$ for $t\leq t_-$ and $t_+$ such that $\Omega_{\textrm{min}}(t) \geq \frac{1}{2}$ for $t\geq t_+$. For $t\geq t_+$, the Gaussian in (\ref{Wigev2}) is wider than a coherent state so $W(\x,t)$ can be interpreted as a Gaussian convolution of the Husimi function, and is therefore positive. On the other hand, for $t\leq t_-$, the Husimi function itself can be interpreted as a convolution of $W(\x,t)$ with a Gaussian, hence, if the Husimi has zero(s), then $W(\x,t)$ must have negative regions. Notice that, although for $N=1$ Hudson theorem \cite{Hudson} ensures that the Husimi of a non Gaussian $W(\x,0)$ necessarily has at least a zero, the argument does not hold for $N\geq 2$, as these zeros cannot be mapped to the ones of the Bargmann function any more \cite{Bargmann}. Instead of isolated zeros, one should picture {\it a priori} a codimension-2 manifold, resulting from the intersection of $\Re\left(\langle\Veta|\psi(0) \rangle\right) = 0$ and $\Im\left(\langle\Veta|\psi(0) \rangle\right) = 0$.

One should note that the positivity bounds $t_{\pm}$ depend only on the eigenvalues of $\M(-t)$,
so that they are independent of the initial state.
Between the two bounds $t_{\pm}$ one can make a rough estimate of a positivity threshold as 
$\det \M(-t_p) = e^{4\gamma t_p} \det\M(t_p) = 4^N$,
which dispenses with the diagonalization of $\M(-t)$. A further note is that the assertion of positivity by comparison
with the Husimi function holds even in the absence of dissipation, without any final equilibrium state.

\section{P positivity}
\label{positivity}

The positivity of the evolved Wigner function is a sure indication of the loss of quantum coherence,
but it does not guarantee that an initially entangled state has eventually achieved classical separability.
For this purpose let us consider a decomposition in terms of generalized coherent states $|\Veta, {\mathbf S} \rangle$,
with their respective Wigner functions expressed as (\ref{Wcoherent}).
Then recalling the Glauber-Sundarshan P representation of the density operator (see, e.~g., Ref. \cite{MandelWolf}), 
its symplectic generalization is defined as 
\be
\hat{\rho} = \int d\Veta ~ {\rm P}_{{\mathbf S}}(\Veta)~ |\Veta, {\mathbf S} \rangle\langle\Veta, {\mathbf S}|.
\label{rhoP}
\ee
If ${\rm P}_{{\mathbf S}}(\Veta)$ is a  positive function of $\Veta$, the density operator (\ref{rhoP}) 
is a probability distribution over the density operators of the generalized coherent states, which are product states
of simple coherent states, each defined on a conjugate plane of the eigenbasis of the matrix ${\mathbf S}^T{\mathbf S}$,
so that $\hat{\rho}$ is separable in this special basis.

Taking the Wigner transform of both sides of (\ref{rhoP}), one obtains 
\begin{eqnarray}
 (\pi\hbar)^N W(\x) =\int d\Veta ~ {\rm P}_{{\mathbf S}}(\Veta)~ 
\exp\left(-\frac{1}{\hbar}(\Veta-\x)\cdot{\mathbf S}^T {\mathbf S}(\Veta-\x)\right),
\label{WigGlau}
\end{eqnarray}
so that the Wigner function is a Gaussian smoothing of the Glauber-Sundarshan P-function.
Just as with the Wigner function, one can now define the P-characteristic function as
\be
\chi_{\rm P}(\Vxi;{\mathbf S}) \equiv 
\int d\Veta~  {\rm P}_{\mathbf S}(\Veta) \exp\left(\frac{i}{\hbar}\Vxi \cdot \J\Veta\right),
\label{Pchar}
\ee
and then the convolution theorem applied to (\ref{WigGlau}) supplies the chord function representing the density operator $\hat\rho$ as
\be
\chi(\Vxi) = \chi_{\rm P}(\Vxi;{\mathbf S}) ~ \exp\left(-\frac{1}{4\hbar}\Vxi\cdot ({\mathbf S}^T {\mathbf S})^{-1}\Vxi\right),
\label{chipchi}
\ee
where we have used $-\J\mS^T\mS\J = \mS^T\mS$.

The condition to have a positive function ${\rm P}_{\mathbf S}(\Veta)$ is therefore the condition that the inverse Fourier transform to (\ref{Pchar}) exists and is positive. This requires that $\chi_{\rm P}(\Vxi;{\mathbf S}) \rightarrow 0$ for $\Vxi\rightarrow \infty$ in all directions, that is, according to Eq. (\ref{chipchi}), and since ${\left({\mathbf S}^T\mathbf S\right)}^{-1}$ is also a positive matrix, that 
\be
\chi(\Vxi)~\exp\left(\frac{1}{4\hbar}\Vxi\cdot ({\mathbf S}^T {\mathbf S})^{-1}\Vxi\right) \rightarrow 0
\label{Pconverge}
\ee
for $\Vxi\rightarrow \infty$ in all directions.
On the other hand, if this condition is satisfied, an explicit form for $\chi(\Vxi)$ then supplies $\chi_{\rm P}(\Vxi;{\mathbf S})$ and its FT leads to ${\rm P}_{\mathbf S}(\Veta)$.

From the evolving chord function (\ref{solutionj}) we obtain the evolving P characteristic function as 
\be
 \chi_{\rm P}(\Vxi;{\mathbf S},t) = \chi(e^{-t\G_\Gamma}\Vxi,0)~ 
\exp \left(-\frac{1}{2\hbar}\Vxi\cdot\Big[\M(t)- ({\mathbf S}^T\mathbf S)^{-1} \Big]~\Vxi\right),
\label{solutionP}
\ee
beyond the time when $\Big[\M(t)- ({\mathbf S}^T\mathbf S)^{-1} \Big]$ becomes a positive matrix.
From then on, the analysis of the P positivity falls back on our previous
treatment of the positivity of the Wigner function, with the sole proviso that
the matrix $({\mathbf S}^T\mathbf S)^{-1}$ is subtracted from $\M(t)$ everywhere.
The various thresholds for positivity depend on the choice of the symplectic frame
determined by $\mathbf S$. If separability is investigated for an experimentally determined 
{\it computational frame} then the full coherent state must be a product in this frame,
i.e. ${\mathbf S}^T {\mathbf S}$ is chosen diagonal in it and the only freedom is in the choice of 
each pair of eigenvalues $\Omega_n$ . 

An essential feature of the characteristic function of a (non-negative) probability density 
is that its modulus is bounded by its value at the origin, proportional to the normalization integral.
In his proof of the central limit theorem \cite{Levy}, Levy conjectures whether this is also a sufficient
property for positivity. Clearly we can now see that this is not so, since it is also a constraint on
pure state Wigner functions, which are not generally positive, yet they also have their own central limit theorem
\cite{TegShap,OzLNP}. On the other hand, this does provide a necessary condition for the positivity of 
the P function: for all $\Vxi\neq 0$, $|\chi_{\rm P}(\Vxi;{\mathbf S},t)| < \chi_{\rm P}(0;{\mathbf S},t)$.  
For the initial state, the explicit bound in terms of the chord function is
\be
(2\pi\hbar)^N |\chi(\Vxi,0)|~ \exp \left(\frac{1}{4\hbar}\Vxi\cdot ({\mathbf S}^T {\mathbf S})^{-1} ~\Vxi\right) < 1.
\ee
It follows that an initial P positive state, that is streched by a metaplectic transformation parametried by $\mathbf S$, will not necessarily continue to be positive, unless expressed in the corresponding basis of likewise streched coherent states.

\section{Evolution of reduced density operators}
\label{evolution}

The {\it partial trace} of two subsystems $B$ and $C$ (with $N = N_B + N_C$) of the evolving 
pure state density operator $\hat \rho$ 
\be
{\hat\rho}_B =  {\rm tr}_C ~\hat{\rho}~~ {\rm and}~~ {\hat\rho}_C ={\rm tr}_B ~\hat{\rho}
\ee 
define the respective {\it reduced density operators}.
If $\hat \rho$ is a pure state, it is known that the {\it purity} of either subsystem is
\be
{\rm tr}_B ~{\hat{\rho}_B}^2 = {\rm tr}_C ~{\hat{\rho}_C}^2 \leq \Big[{\rm tr} ~{\hat{\rho}}^2 = 1\Big],
\ee
so that the {\it linear entropy}
\be
E_l \equiv 1 - {\rm tr}_B ~{\hat{\rho}_B}^2 = 1 - {\rm tr}_C ~{\hat{\rho}_C}^2
\ee
can be adopted as a quantifier of entanglement. 

Recalling that the general relation for the partial trace of an arbitrary operator $\hat O$ in the Weyl representation is the projection
\be
O_B(\x_B) = \int d\x_C ~ O(\x_B,\x_C),
\ee
whereas in the chord representation one merely needs the section
\be
O_B(\Vxi_B) = (2\pi\hbar)^{N_C} \;O(\Vxi_B,\Vxi_C=0),
\ee
then the special normalization of the Wigner function and the chord function leads to the representations 
of the respective reduced density operators as
\be
 W_B(\x_B) = \int d\x_C ~W(\x_B,\x_C) ~~~{\rm and} ~~~ \chi_B (\Vxi_B) = (2\pi\hbar)^{N_C}\;\chi(\Vxi_B, \Vxi_C =0) ~.
\label{reducedW}
\ee
Recalling the general expressions for the trace of the square of an operator \cite{Report}, the linear entropy in these representations becomes
\be
E_l = (2\pi\hbar)^{N_B}\int d\x_B ~[W_B(\x_B)]^2 =  (2\pi\hbar)^{N_B}\int d\Vxi_B ~|\chi_B(\Vxi_B)|^2 ~.
\ee

The reduced density operator contains all the information that can be extracted from any measurement
effected on either subsystem. For instance, for an observable ${\hat O} = {\hat O}_B \otimes{\hat \Id}_C$
\be
\langle \hat{O} \rangle = {\rm tr}~\hat{\rho}~\hat{O} =  {\rm tr}_B~\hat{\rho}_B~\hat{O}_B = \langle \hat{O}_B \rangle ~ .
\label{averager}
\ee 
But given the singular chord representation of the identity operator (\ref{Id}), this equality follows immediately by inserting
this reduced observable into the integral for the chord expectation (\ref{average}), giving back the reduced chord function (\ref{reducedW}). In the case of polynomial functions of positions and momenta defined on subsystem $B$, their evolving expectation can be obtained from Eqs. (\ref{moments}) applied directly to the reduced chord function (\ref{reducedW}).

The reduced chord function for the Markovian evolution of the density operator can be factored in a similar way to the full chord function as
\be
\chi_B(\Vxi_B,t) = \chi_B(\Vxi_B,0) ~ \exp \left(-\frac{1}{2\hbar}\Vxi_B \cdot \M_B(t) ~ \Vxi_B   \right)~,
\label{solutionB}
\ee
with the definitions of the reduced quadratic form
\be
\Vxi_B \cdot \M_B(t) ~ \Vxi_B \equiv (\Vxi_B, \Vxi_C =0)\cdot \M(t)~(\Vxi_B, \Vxi_C =0)
\ee
and 
\be
\chi_B(\Vxi_B,t) \equiv (2\pi\hbar)^{N_C} \chi\left(e^{-t\G_\Gamma}(\Vxi_B,\Vxi_C=0),0\right) ~.
\ee
Even though the linear classical evolution of the chords will rotate the $\Vxi_C=0$ plane  
in the full $(2N)$ D phase space and there generally will be dissipation, $\chi_0(\Vxi_B,t)$ is correctly normalized
at the origin as a {\it reduced decoherentless chord function}. 

Inserting the reduced chord function (\ref{solutionB}) into the Fourier expression for the reduced Wigner function
\be
W_B(\x_B,t) = \int \frac{\mathrm{d}{\Vxi_B}}{(2\pi\hbar)^{N_B}}~ \chi_B(\Vxi_B,t)~ \exp \left[\frac{i}{\hbar}\x_B \cdot \J\Vxi_B \right] ~,
\label{FourierWignerr}
\ee
and introducing (\ref{solutionB}) leads to 
\begin{eqnarray}
 W_B(\x_B,t) &=& \frac{1}{\sqrt{\det\M_B(t)}} \int \frac{d\x'_B}{(2\pi\hbar)^{N_B}}~ \W_{B}(\x_B',0) \label{Wigev2b} \\   \nonumber
& & \exp\left[-\frac{1}{2\hbar} (\x_B'-\x_B) \cdot {\M_\J}_B(-t)^{-1}(\x_B'-\x_B)\right] ~,
\end{eqnarray}
with ${\M_\J}_B(-t) = -\J_B \M_B (-t)\J_B$.
Notwithstanding the less transparent classical evolution supporting $\W_{B}(\x_B',0)$  
than for the full evolving Wigner function, we still retrieve the effect on the reduced Wigner function 
of the Markovian coupling to the environment as a convolution to a widening Gaussian window. Of course,
the reduced decoherentless Wigner function is not itself a pure state in general. So one may expect that
positivity of reduced Wigner functions may well precede that of the Wigner function in the full $(2N)$ D phase space.  

\begin{figure}[!ht]
\begin{center}
\includegraphics[width=16cm]{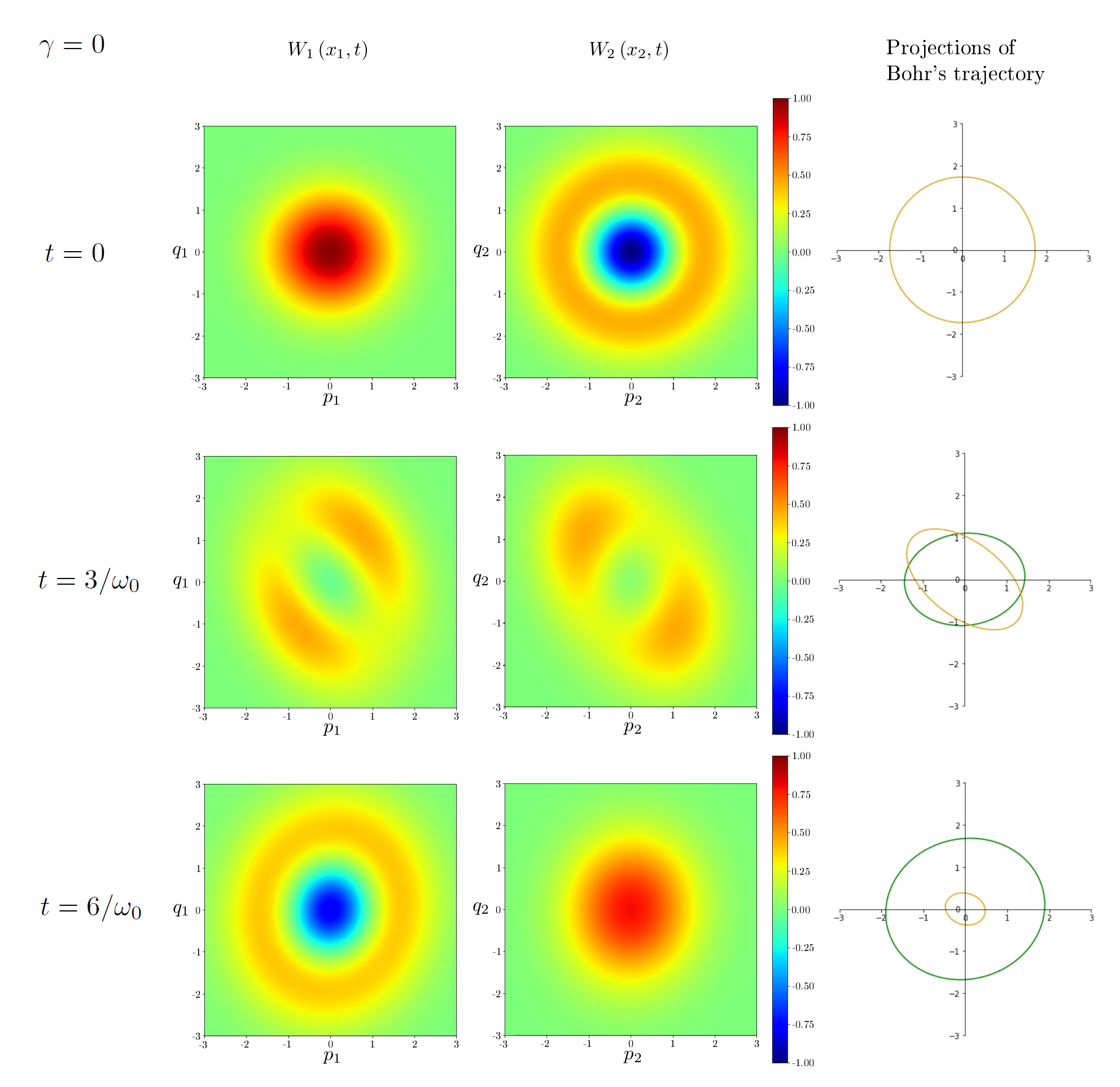}
\caption{Evolution of the reduced Wigner functions of the state of two coupled harmonic oscillators without interaction with the environment, that is, $\gamma=0$, at three different instants of time. The first column shows the reduced Wigner function of the $\x_1$ oscillator, which is initially in state $n_1=0$. The second column shows the Wigner function of the $\x_2$ oscillator, which is initially in state $n_2=1$. Then the third column shows the evolution of the projection of the corresponding Bohr orbit in each phase space. Initially this orbit is a circle in the $\x_2$ space, with no extension in the $\x_1$ space, thus coinciding with its orange $\x_2$ projection. Its radius is set to $\sqrt{3\hbar}$, with $\hbar=1$, so as to get the quantized area $h\left(n_2+\frac{1}{2}\right)$. The Hamiltonian dynamics then turns this circle into an ellipse in the whole $(\x_1,\x_2)$ phase space, so its $\x_1$ green projection grows while its $\x_2$ orange one shrinks.}
\label{RedWignerg0}
\end{center}
\end{figure}
\begin{figure}[!ht]
\begin{center}
\includegraphics[width=16cm]{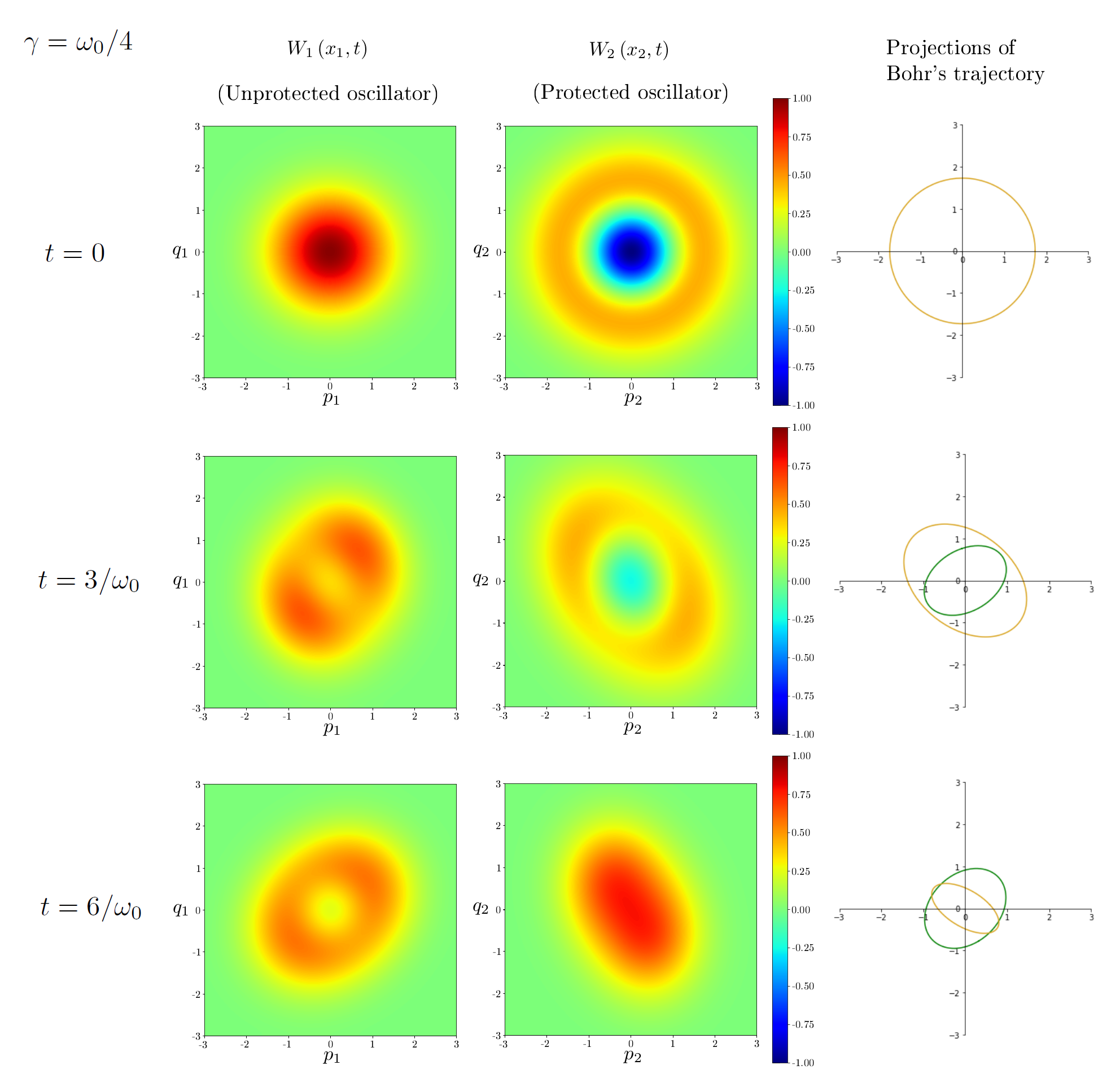}
\caption{Represented features and the context are the same as in Fig. \ref{RedWignerg0}, except that we set $\gamma=\omega_0/4$, so the first oscillator, with initial state $n_1=0$, is now ``unprotected''. We still can see the excited state and the Bohr orbit switching alternatively from one oscillator to the other, but, simultaneously, the Wigner functions get damped, and the corresponding Bohr orbit globally shrinks, which can be seen from both its projections.}
\label{RedWigner}
\end{center}
\end{figure}

As an example,  Fig. \ref{RedWignerg0} and Fig. \ref{RedWigner} show the plots of the reduced Wigner functions for two harmonic oscillators with the same natural frequency $\omega_{0}$ and which are coupled through their variables by a term $c p_{1} p_{2}$, where we use $c = \omega_{0}/2$. One of the oscillators, called ``unprotected,'' is coupled to the environment with a strength $\gamma$, whereas the other one is ``protected,'' that is, without interaction with the environment. The details of the calculation are presented in the following section and in Appendix \ref{triat}. The initial state is a product of the fundamental state $n=0$ for the unprotected oscillator and the first excited state $n=1$ for the protected oscillator. Being protected, the latter would never become positive, were it not coupled to the unprotected oscillator. Because of this coupling, the whole state reaches positivity. The mechanism of this indirect decoherence is a cyclic beating between both oscillators, where each state of the initial product alternatively spends some time under the unprotected regime.

Three instants of time are considered, $t=0$, $t=3/ \omega_{0}$ and $t=6/ \omega_{0}$. Fig. \ref{RedWignerg0} illustrates the ``pure beating'' case $\gamma=0$, that is, when both oscillators are protected from the environment, and we can see the $n=1$ state being transfered from one oscillator to the other. On the other hand, Fig. \ref{RedWigner} illustrates the decoherence induced by chosing for instance $\gamma = \omega_{0}/4$, thus unprotecting one of the oscillators. Although the beating is still visible, both reduced Wigner functions will eventually become fully positive, illustrating the positivity effect contained in (\ref{Wigev2b}).

The chosen initial state corresponds semiclassically to Bohr's first excited trajectory in the phase space of the protected oscillator. Then its time evolution is also shown, in order to emphasize its relevance to the form of reduced Wigner functions. Its projection on its initial protected plane, is shown in orange, whereas its projection on the unprotected plane is shown in green.

\section{Protection from dissipation}
\label{protect}

A system with one degree of freedom is either isolated or it interacts more or less strongly with the environment.
In contrast, in a system with several degrees of freedom, some of the variables may be relatively protected from
direct external forces, while their influence is predominantly transmitted by the internal coupling 
among the degrees of freedom. It is this richer scenario that we first choose to exemplify the complexity
of the Lindbladian evolution when there is more than one degree of freedom. For simplicity, we limit the analysis to a pair of degrees of freedom and to the extreme case of Lindblad operators acting only on one of them. 
In Appendix \ref{triat} we sketch the derivation of the Hamiltonian for a symmetric triatomic molecule, with its symmetry
broken only by an isotopic mass difference between the external atoms. Within a symplectic transformation, the approximate Hamiltonian  (\ref{H2}) becomes
\begin{equation}
H\left(\x\right)=\frac{\omega_{1}}{2}\left(p_{1}^{2}+q_{1}^{2}\right)+\frac{\omega_{2}}{2}\left(p_{2}^{2}+q_{2}^{2}\right)+c~p_{1}p_{2} ~.
\label{H2app}
\end{equation}
As a practical example, one can think about a CO$_2$ molecule with isotopes $^{16}$O and $^{18}$O. As it is shown in Appendix \ref{triat}, one then has $\omega_1 \simeq  6.43\times 10^{14}$ rad.s$^{-1}$, $\omega_2 \simeq  5.35\times 10^{14}$ rad.s$^{-1}$ and $c \simeq  0.57\times 10^{14}$ rad.s$^{-1}$. The resulting condition $c/\omega_1\simeq c/\omega_2\simeq 1/10$ motivates the following treatment as a perturbative expansion in powers of $c$.

The main interaction with the radiation in the environment is mediated by the oscillating dipole moment of the fully asymmetric mode, whereas the symmetric one, as a quadripole, is comparatively protected. The scheme can be understood from Fig. \ref{coord}, where the barycenter of the two negative lateral atoms and of the positive central one is only modified by an asymmetric oscillation. Thus we here postulate the single Lindblad operator $\hat L = \sqrt{\frac{\gamma}{2}} ~ {\hat a_1}$ with the Weyl representation
\be
a_1(\x) = \frac{q_1 + i~p_1}{\sqrt{2}} ,
\ee 
so that the dissipation coefficient (\ref{dissipcoef}) is just $\gamma$ and the classical evolution matrix (\ref{defG}) becomes
\begin{equation}
\G_\Gamma = \J\mH+\GGamma=\left[\begin{array}{cccc}
\gamma & -\omega_{1} & 0 & 0\\
\omega_{1} & \gamma & c & 0 \\
 0 & 0 & 0 & -\omega_{2}\\
c & 0 & \omega_{2} & 0
\end{array}\right] ~.
\end{equation}
The dependence of the eigenvalues on the isotopic parameter is of second order with respect to the fully symmetrical system, which decouples as
\be
\lambda_{1\pm}^{(0)}=\gamma\pm i\omega_{1} ~~~~~~  \lambda_{2\pm}^{(0)}=\pm i\omega_{2} ~.
\label{eigenv}
\ee
On the other hand, the first order expansion of the complex eigenvectors is
 \begin{equation}
V_{i\pm}=V_{i\pm}^{\left(0\right)}+cV_{i\pm}^{\left(1\right)} ~,
\end{equation}
where the uncoupled eigenvectors are simply
\begin{equation}
\begin{array}{cc}
V_{1\pm}^{\left(0\right)}=\frac{1}{\sqrt{2}}\left(\begin{array}{c}
\pm i\\
1\\
0\\
0
\end{array}\right) & ~~~~ V_{2\pm}^{\left(0\right)}=\frac{1}{\sqrt{2}}\left(\begin{array}{c}
0\\
0\\
\pm i\\
1
\end{array}\right)\end{array}  ~.
\end{equation}
The first order perturbation of the eigenmodes can then be expressed as
\begin{equation}
\begin{array}{cc}
V_{1\pm}^{\left(1\right)}=\frac{e^{\pm i\phi_{1}}}{\sqrt{2}\rho_{1}}\left(\begin{array}{c}
0\\
0\\
-\omega_{2}\\
\varrho_{1}e^{\pm i\varphi_{1}}
\end{array}\right) & ~~~~  V_{2\pm}^{\left(1\right)}=\frac{e^{\pm i\phi_{2}}}{\sqrt{2}\rho_{2}}\left(\begin{array}{c}
-\omega_{1}\\
\varrho_{2}e^{\pm i\varphi_{2}}\\
0\\
0
\end{array}\right)\end{array} ~,
\end{equation}
with the suplementary definitions
\begin{equation}
\begin{array}{ll}
\rho_{1}=\sqrt{\left(\omega_{2}^{2}-\omega_{1}^{2}+\gamma^{2}\right)^{2}+\left(2\omega_{1}\gamma\right)^{2}} & ~~~ \varrho_{1}=\left[\omega_{1}^{2}+\gamma^{2}\right]^{\frac{1}{2}}\\
\\
\phi_{1}=tan^{-1}\left(2\omega_{1}\gamma/\left(\omega_{2}^{2}-\omega_{1}^{2}+\gamma^{2}\right)\right)& ~~~ \varphi_{1}=tan^{-1}\left(\omega_{1}/\gamma\right)\\
\\
\rho_{2}=\sqrt{\left(\omega_{2}^{2}-\omega_{1}^{2}-\gamma^{2}\right)^{2}+\left(2\omega_{2}\gamma\right)^{2}} & ~~~ \varrho_{2}=\left[\omega_{2}^{2}+\gamma^{2}\right]^{\frac{1}{2}}\\
\\
\phi_{2}=tan^{-1}\left(2\omega_{2}\gamma/\left(\omega_{2}^{2}-\omega_{1}^{2}-\gamma^{2}\right)\right)  & ~~~ \varphi_{2}=-tan^{-1}\left(\omega_{2}/\gamma\right) ~,
\end{array}
\end{equation}
(Note that only the last relation has a negative sign.) 

Superposing complex eigenvectors
\be
V_{ie} = \frac{V_{i+} + V_{i-}}{2} ~~~{\rm  and} ~~~ V_{io} = \frac{V_{i+} - V_{i-}}{2i}
\ee
defines a pair of real phase planes, the real eigenmodes, not coincident with the symmetrical and antisymmetrical planes coordinated by $\x_i$.
Let us consider a trajectory initially in the symmetrical $\x_1=0$ plane which is protected from dissipation. Its main projection is in the 
$\x'_1 = 0$ eigenplane (spanned by $V_{2e}$ and $V_{2o}$), but also a small projection on the other eigenplane; see Fig. \ref{RedWignerg0} and Fig. \ref{RedWigner}. According to (\ref{eigenv}),
to first order in $c$, there will be pure rotation in the $\x'_1= 0$ plane. The interesting point is that the small component in the other
real eigenplane spirals outwards exponentially. Thus, if we associate this choice of initial value to the chord $\Vxi$ in the argument of the
evolving chord function in (\ref{solutionj}), the backward propagation of $\Vxi$ will reach $\Vxi=0$ exponentially fast in (\ref{defM}), leading to a fast convergence of the integrand (\ref{Ct}) to a finite value in the exponential.

\section{Network of strongly coupled harmonic oscillators with dissipation on its surface}
\label{network}

We study a linear chain of $N$ harmonic oscillators coupled by
$\hbar\alpha\frac{\oper{a_n}^\dagger\oper{a_{n+1}}+\oper{a_{n+1}}^\dagger\oper{a_n}}{2}$, with
\begin{eqnarray}
\oper{a_n} = \sqrt{\frac{m\omega}{2\hbar}}\oper{q_n} + i \frac{1}{\sqrt{2\hbar m\omega}}\oper{p_n} \cr
\oper{a_n}^\dagger = \sqrt{\frac{m\omega}{2\hbar}}\oper{q_n} - i \frac{1}{\sqrt{2\hbar m\omega}}\oper{p_n}.
\end{eqnarray}
 The Hamiltonian is
\be
\oper H = \sum_{n=1}^N \left( \frac{\oper{p_n}^2}{2m} + \frac{1}{2}m\omega^2 \oper{q_n}^2 \right)
+\sum_{n=1}^{N-1} \left(\frac{m\omega\alpha}{2} \oper{q_n}\oper{q_{n+1}} + \frac{\alpha}{2 m\omega} \oper{p_n}\oper{p_{n+1}}\right).
\ee

On top of this, we assume that the oscillator no. 1 is coupled to a Markovian environment, whereas the coupling of other oscillators to environment is supposed to be negligible. An experimental realization of this system is a network of photon cavities coupled with each other, while the one on the edge of the network is coupled to the environment. This can be modeled through a Lindblad equation
\be
 \frac{d\oper\rho}{dt} = -\frac{i}{\hbar}\left[\oper H,\oper\rho\right] - \frac{\gamma \bar n}{2}\left[ 2\oper {a_1}^\dagger \oper \rho\oper{a_1} - \oper {a_1}\oper {a_1}^\dagger\oper \rho - \oper \rho \oper {a_1}\oper {a_1}^\dagger\right]
- \frac{\gamma (\bar n+1)}{2}\left[ 2\oper {a_1} \oper \rho\oper{a_1}^\dagger - \oper {a_1}^\dagger\oper {a_1}\oper \rho - \oper \rho \oper {a_1}^\dagger\oper {a_1}\right].
\label{lind}
\ee
The equation fulfills detailed balance. For that, departing from the simplification in the previous section, we need two linear Lindblad operators, $\oper L_e = \oper a_1^\dagger = \vct l_e\cdot \oper{\vct x}$ and $\oper L_d = \oper a_1 = \vct l_d\cdot \oper{\vct x}$, defining two complex vectors $\vct l_e=\vct l_e' + i\vct l_e''$ and $\vct l_d=\vct l_d' + i\vct l_d''$ with
\be
\vct l_e' = \left(\begin{array}{c} 0 \cr \sqrt{\frac{m\omega\gamma  \bar n}{2}} \cr 0 \cr \vdots \end{array}\right) ~~~
\vct l_e'' = \left(\begin{array}{c}  -\sqrt{\frac{\gamma  \bar n}{2m\omega}} \cr 0 \cr 0 \cr \vdots  \end{array}\right) ~~~
\vct l_d' = \left(\begin{array}{c} 0 \cr \sqrt{\frac{m\omega\gamma  (\bar n+1)}{2}} \cr 0 \cr \vdots  \end{array}\right) ~~~
\vct l_d'' = \left(\begin{array}{c} \sqrt{\frac{\gamma  (\bar n+1)}{2m\omega}} \cr 0 \cr 0 \cr \vdots  \end{array}\right).
\ee
Still, because of the balance condition, the decoherence rate matrix $\matr \Gamma$ is diagonal, 
\be
\matr \Gamma = \matr \Gamma_e + \matr \Gamma_d = \mD_{\frac{\gamma}{2},\frac{\gamma}{2},0,\ldots,0}.
\ee
 The aim of this section is to describe the spectrum $\left\{\lambda_{k\pm}\right\}_{k=1,\ldots,N}$ of the classical propagation matrix $\G_\Gamma$, defined by (\ref{defG}),  at first order in $\frac{\gamma}{\alpha}$, that is, for small coupling to the environment as compared to the internal coupling. 

As it is shown in Appendix \ref{eigen}, for $\gamma=0$, $\G_\Gamma$ is the product of a T\oe plitz matrix with a 1D harmonic oscillator, and it has the exact eigenvalues
\be
\lambda_{k\pm}^{(0)} = \pm i\left( \omega + \alpha \cos{\left(\frac{\pi k}{N+1}\right)} \right).
\ee
Then, for small $\frac{\gamma}{\alpha}$, the eigenvalues of $\G_\Gamma$ can be computed perturbatively, giving, at first order
\be
\lambda_{k\pm}^{(1)} = \pm i\left( \omega + \alpha \cos{\left(\frac{\pi k}{N+1}\right)} \right) + \frac{\gamma}{2(N+1)} \sin^2{\left(\frac{k\pi}{N+1}\right)} .
\label{eigenG}
\ee
Notice that their real part, corresponding to the dissipation rate, goes as $\frac{1}{N^3}$ for the frequencies that are close to $\omega\pm\alpha$, corresponding to fixed $k$ or fixed $N+1-k$ as $N\rightarrow\infty$. We call these frequencies ``band-edge modes'', as they are on the edge of the frequency interval. On the other hand, dissipation goes as $\frac{1}{N}$ for frequencies close to $\omega$, corresponding to fixed $k- \frac{N}{2}$ as $N\rightarrow\infty$. We call these frequencies ``band-center modes'' -- see Fig. \ref{mod}. 

The eigenvectors $\vct W_{k\pm}$ of $\G_\Gamma$ are calculated in Appendix \ref{eigen} and will serve as a basis to decompose generic chords, which are real even though the chord function is complex. To restrict the chord function to the $k^{\textrm{th}}$ mode one takes $\Vxi$ of the form $\Vxi_k=z\vct W_{k+}+z^*\vct W_{k-}$, with $z\in{\mathbb C}$ and where $\vct W_{k+}=\vct W_{k-}^*$ are defined in (\ref{defW}). Having a real $\Vxi_k$ then implies that $z=\sqrt{m\omega+\frac{1}{m\omega}}\left(\frac{\xi_p}{2\sqrt{m\omega}}+i\frac{\sqrt{m\omega}}{2}\xi_q\right)$, which expands like
\be
 \Vxi_k = \vct V_{k}^{(0)}\otimes (\xi_p,\xi_q)^T + \GO\left(\left(\frac{\gamma}{\alpha}\right)^2\right),
\ee
with $(\xi_p,\xi_q)\in\mathbb R^2$ and $V_{k}^{(0)}$ defined by (\ref{Vk0}). The picture of the $k^{\textrm{th}}$ mode is then $N$ sites undergoing the same $(\xi_p,\xi_q)$ oscillation, but with different amplitudes -- see FIG. \ref{mod}. Then, according to (\ref{solutionj}), (\ref{explicitM}) and (\ref{Ct}), one has, at first order in $\frac{\gamma}{\alpha}$,
\begin{eqnarray}
 \chi\left(\Vxi_k,t\right) &\simeq& \chi\left(\vct V_{k}^{(0)}\otimes\left( z e^{-\lambda_{k+}t}\epsilon_+ + z^* e^{-\lambda_{k-}t}\epsilon_- \right),0\right)~ \\ \nonumber 
  & & \exp \left[-\frac{1}{2\hbar}
\left(2\bar n+1\right)\left(1-e^{-\frac{\gamma}{N+1}\sin^2{\left(\frac{k\pi}{N+1}\right)}t}\right)\left(\frac{\xi_p^2}{m\omega} + m\omega\xi_q^2\right)
 \right]~,
\end{eqnarray}
with the final equilibrium state,
\be
 \chi\left(\Vxi_k,+\infty\right) \simeq \left(2\pi\hbar\right)^{-N}\exp \left[-\frac{1}{2\hbar}
\left(2\bar n+1\right)\left(\frac{\xi_p^2}{m\omega} + m\omega\xi_q^2\right)
 \right]~,
\label{final}
\ee
where $2\bar n+1 = \frac{1}{\tanh{\left(\frac{\beta\hbar\omega}{2}\right)}}$, so we recognize the chord representation of a thermal state.
\begin{figure}[!ht]
\begin{center}
\includegraphics[width=14cm]{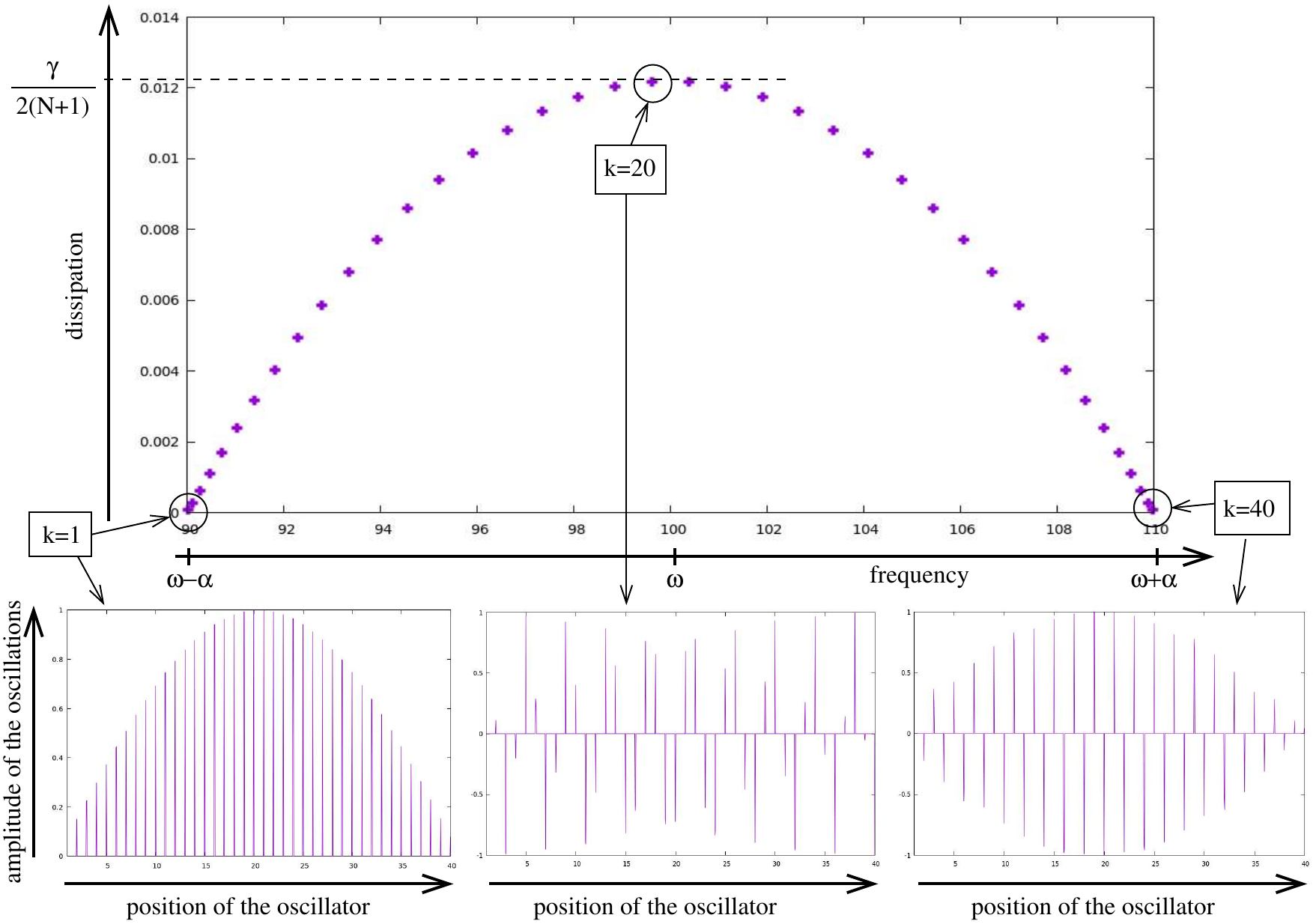}
\caption{Upper graph shows the eigenmodes (\ref{eigenG}) of $\G_\Gamma$ for $N=40$, with the imaginary part (frequency) on the horizontal axis and the real part (dissipation) on the vertical axis. The modes $k=1$ and $k=40$ are called ``band-edge modes'' because they are on the edges of the frequency interval. They are represented in the lower left and right graphs, which show the $(\xi_p,\xi_q)$ amplitude as a function of the position of the oscillator in the network. These modes are the most slowly affected by dissipation, that is, their dissipation rate is of order $\gamma/N^3$, which can be related to the fact that their amplitude is small close to the surface of the network.
Mode $k=20$ is a ``band-center mode'' because it is in the center of the frequency interval. It is represented in the lower middle graph. It is the mode which is the most rapidly affected by dissipation, that is, its dissipation rate is of order $\gamma/N$. It can be related to the fact that it has a nonvanishing amplitude close to the surface of the network.}
\label{mod}
\end{center}
\end{figure}
From these expressions we can deduce two things. First, that band-edge modes converge to their final equilibrium in a time of order $N^3/\gamma$,  whereas band-center modes converge in a time $N/\gamma$. Secondly, that each mode converges to the same thermal mode (\ref{final}). This may seem surprising, as the terms $1/(\lambda_{j\pm}+\lambda_{k\pm})$ in (\ref{Cinfini}) seem to give more weight to the band-edge modes, whose real part of the eigenvalues goes as $1/N^3$. However, this weight is actually compensated through (\ref{B}) by the fact that band-edge modes have smaller amplitude $(\vl'\cdot\vct W_{1\pm})^2\simeq \frac{\gamma}{N^3}$ at the border of the network, where the Lindblad operators act. On the other hand, while the band-center modes have the real part of the eigenvalues in $1/N$, giving a smaller contribution in (\ref{Cinfini}), their amplitude at the border, $(\vl'\cdot\vct W_{\frac{N}{2}\pm})^2\simeq \frac{\gamma}{N}$, is larger -- see (\ref{Vk0}) in Appendix \ref{eigen}. Notice that the present perturbation regime $\gamma\ll\alpha$ is the inverse of the one of Sec. \ref{protect}, so one does not expect a protected plane here.

Instead of looking at the $k^{\textrm{th}}$ mode, one can also look at the reduced chord function of the $n^{\textrm{th}}$ harmonic oscillator, which, according to (\ref{reducedW}), is 
\be
\chi_n\left( \xi_p , \xi_q ,t=+\infty\right) = \chi\left( \mathop{\vct 0}_1, \ldots,\mathop{\vct 0}_{n-1},(\xi_p,\xi_q),\mathop{\vct 0}_{n+1},\ldots,\mathop{\vct 0}_N ,t=+\infty\right).
\ee
To obtain its expression, we decompose the vector $\left(0,\ldots,0,1,0,\ldots,0\right)^T\otimes(\xi_p,\xi_q)^T$ into the eigenbasis $\vct W_{k\pm}$, thus obtaining the $\mP^{-1}\vct\xi$ of (\ref{solutionfty}), and the corresponding final state
\be
\chi_n\left( \xi_p , \xi_q ,+\infty\right) = \left(2\pi\hbar\right)^{-N}\exp \left[-\frac{1}{2\hbar}
\left(2\bar n+1\right)\left(\frac{\xi_p^2}{m\omega} + m\omega\xi_q^2\right) \Sigma_N \right]
\label{finalnth}
\ee
with
\be
  \Sigma_N =  \frac{4}{(N+1)^3}
\sum_{j=1}^N\sum_{k=1}^N 
\frac{
\sin{\frac{j\pi}{N+1}}\sin{\frac{nj\pi}{N+1}}\sin{\frac{k\pi}{N+1}}\sin{\frac{nk\pi}{N+1}}
\left(\sin^2{\left(\frac{j\pi}{N+1}\right)}+\sin^2{\left(\frac{k\pi}{N+1}\right)}\right)
}{
\left(\frac{\sin^2{\left(\frac{j\pi}{N+1}\right)}+\sin^2{\left(\frac{k\pi}{N+1}\right)}}{N+1}\right)^2
+\frac{4\alpha^2}{\gamma^2}\left(\cos{\frac{j\pi}{N+1}}-\cos{\frac{k\pi}{N+1}}\right)^2
}.
\label{sigmaN}
\ee
The expression (\ref{sigmaN}) looks fairly heavy, but the condition $\gamma\ll\alpha$ ensures that the double sum will be dominated by its diagonal $j=k$, which shows that actually $\Sigma_N\simeq 1$ and
\be
\chi_n\left( \xi_p , \xi_q ,+\infty\right) \simeq \left(2\pi\hbar\right)^{-N}\exp \left[-\frac{1}{2\hbar}
\left(2\bar n+1\right)\left(\frac{\xi_p^2}{m\omega} + m\omega\xi_q^2\right)\right]~.
\label{finalnthsimp}
\ee
Hence, the $n^{\textrm{th}}$ harmonic oscillator also converges towards the thermal state, on account of its coupling to the environment through its neighbours.

\begin{figure}[!ht]
\begin{center}
\includegraphics[width=8cm]{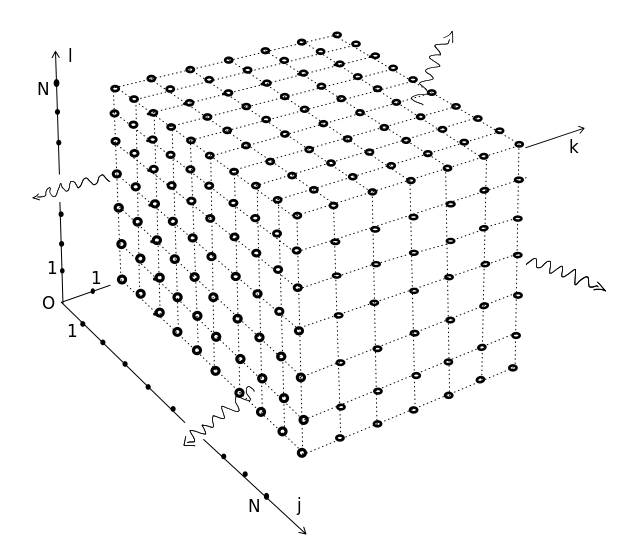}
\caption{Cubic network of harmonic oscillators undergoing dissipation on its surface, but not inside. }
\label{cub}
\end{center}
\end{figure}
The result can be generalized to a cubic network of harmonic oscillators, see Fig. \ref{cub}, 
with annihilation and creation operators $a_{j,k,l}$ and $a_{j,k,l}^\dagger$. Its surface, corresponding to $j,k,l\in\left\{1,N\right\}$, undergoes dissipation through a Lindblad equation of type (\ref{lind}). Then the eigenvalues of the evolution matrix $\G_\Gamma$ are, at first order in $\gamma/\alpha$,
\begin{eqnarray}
 \lambda_{n_x,n_y,n_z\pm}^{(1)} = \pm i\left[ \omega + \alpha \left(\cos{\left(\frac{\pi n_x}{N+1}\right)} + \cos{\left(\frac{\pi n_y}{N+1}\right)} + \cos{\left(\frac{\pi n_z}{N+1}\right)}\right)\right] \nonumber\\ 
 + \frac{\gamma}{N+1} \left( \sin^2{\left(\frac{\pi n_x}{N+1}\right)} +\sin^2{\left(\frac{\pi n_y}{N+1}\right)}+\sin^2{\left(\frac{\pi n_z}{N+1}\right)} \right), 
\end{eqnarray}
while the eigenvectors at order zero are just the tensor products of three vectors of (\ref{Vk0}) type.

\section{Conclusion}
In this article we derived the Wigner function of a state driven by a quadratic Hamiltonian linearly coupled to a Markovian environment. It can be written as the convolution of a broadening Gaussian with the Liouville propagation of the initial Wigner function. This expression is derived from the simpler expression of its Fourier transform, the chord function, which is the product of a shrinking Gaussian with the Liouville propagation of the initial chord function. Thus, the latter can be interpreted as the characteristic function of a signal getting through a Gaussian channel, whose illustrative example could be an idealized optical fiber. The dynamics of the Gaussian depends on the real part of the eigenvalues of the matrix $\G_\Gamma$ driving the Liouville propagation. If all the real parts are positive, then the Wigner function converges to a finite positive Gaussian. If some real parts are negative, then the final Gaussian has infinite width in some directions.

The Wigner function becomes positive at the instant at which the convoluting Gaussian reaches the size of a coherent state in every direction of phase space. This instant is determined by the real part of the eigenvalues of the matrix $\G_\Gamma$. In the general case with several degrees of freedom, the Gaussian will not reach this size at the same moment for each of the corresponding eigenspaces, so this defines a time range $[t_{\textrm{min}},t_{\textrm{max}}]$ during which the restriction of the Wigner function becomes positive on a growing subspace. Full positivity is only guaranteed for $t\geq t_{\textrm{max}}$. On the other hand, for an initial state which is not a Gaussian, the existence of negative parts for $t<t_{\textrm{min}}$ is only certain for one degree of freedom, because the Husimi function then maps with the Bargmann function.

The chord function of the system is especially convenient as it gives easy access to the chord symbol of any reduced density operator obtained from tracing out a subsystem. One just has to set to zero all the variables corresponding to the subsystem that is traced out. This allows one to derive simple expressions for the linear entropy.

When each eigenmode of the quadratic Hamiltonian is independently coupled to the environment, then the global chord function is just a product of one degree of freedom chord functions. More complex behaviours are expected when degrees of freedom which are coupled to environment do not coincide with eigenmodes of the Hamiltonian. Then two examples show how internal degrees of freedom which are not directly coupled to the environment can undergo decoherence through internal coupling of the system. Another point of view is that one can predict the sensitivity of an eigenmode to decoherence by evaluating its overlap with the degrees of freedom which are coupled with the environment. If this overlap is zero then the eigenmode is protected.

\appendix

\section{Broken symmetry of a linear triatomic molecule}
\label{triat}

Consider a linear triatomic molecule that is symmetric as far as charges are concerned, but with different isotopes for the end atoms.
\begin{figure}[!ht]
\begin{center}
\includegraphics[width=12cm]{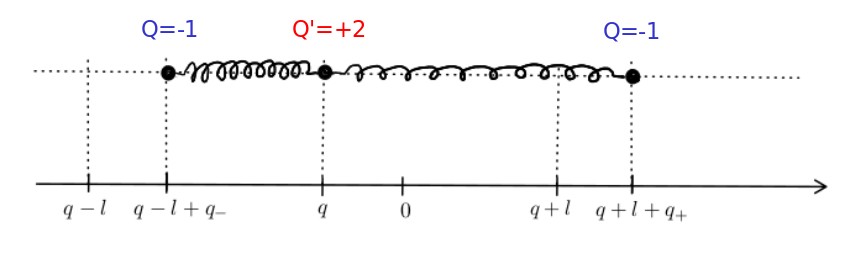}
\caption{Triatomic molecule with convenient choice of coordinates. Lateral atoms are negatively polarized whereas the central one is positive.}
\label{coord}
\end{center}
\end{figure}
Choosing the coordinates as in the Fig. \ref{coord}, the equilibrium lengths take no part in the Lagrangian:
\be
\textsl{L}_0 = \frac{m}{2}~{\dot q}^2 + \frac{m_+}{2}~(\dot q+{\dot q}_+)^2 + \frac{m_-}{2}~(\dot q+{\dot q}_-)^2 - \frac{k}{2}~({q_-}^2 +{q_+}^2)~.
\ee
Conservation of total momentum
\be
P =m~{\dot  q} + m_+~({\dot q} + {\dot q}_+) + m_-~({\dot q} + {\dot q}_-) 
\ee
together with the definitions
\be
M=m + m_+ + m_- ~~~ {\rm and} ~~~ \mu_\pm = \frac{m_\pm}{M}
\ee
lead to 
\be
{\dot q} = - \mu_+ {\dot q}_+ - \mu_- {\dot q}_-
\ee
and to the reduced Lagrangian
\be
 L_1 = \frac{M}{2}~\left[(\mu_+ - \mu_+^2)~ {\dot q}_+^2 + (\mu_- - \mu_-^2)~ {\dot q}_-^2 
      -2 \mu_+\mu_- ~{\dot q}_+ {\dot q}_- - ~\omega_0^2~({q_-}^2 +{q_+}^2)\right]~,
\ee
with $\omega_0^2 = k/M$. 

The new variables
\begin{eqnarray}
 q_1 = \frac{q_+ + q_-}{\sqrt{2}} ~~~~~~  q_2 = \frac{q_+ - q_-}{\sqrt{2}}
\end{eqnarray}
represent pure symmetric streching for $q_1 = 0$, whereas $q_2 =0$ denotes equal displacement of both end atoms with respect to the central atom,
i.e. antisymmetrical. Discarding the factor $M/2$ and further defining
\begin{eqnarray}
a = (\mu_+ - \mu_+^2) + (\mu_- - \mu_-^2)- 2\mu_+ \mu_-  \\   \nonumber
b = (\mu_+ - \mu_+^2) + (\mu_- - \mu_-^2)+ 2\mu_+ \mu_-   \\
\epsilon = (\mu_+ - \mu_+^2 - \mu_- + \mu_-^2) ~,
\end{eqnarray}
the Lagrangian in the new variables takes the form
\begin{eqnarray}
L_2 = \frac{a}{2} ~{\dot q}_1^2 
+ \frac{b}{2} ~{\dot q}_2^2  
+ \epsilon ~ {\dot q}_1{\dot q}_2
- ~\omega_0^2~({q_1}^2 +{q_2}^2) ~,
\end{eqnarray} 
which in the case of equal end-masses, such that $\mu_+ = \mu_- = \mu$, is separable:
\be
L_s = \mu ~ {\dot q} _2^2 - \omega_0^2 ~ q_2^2+ (\mu - 2\mu^2)~ {\dot q_1}^2 - \omega_0^2 ~q_1^2 ~. 
\ee

For small coupling due to isotopic variation the normal modes of $L_2$ will not coincide exactly with the symmetric and anti symmetric variables.
Rather than deriving here their explicit, though not very transparent form, it is more relevant in the ulterior quantum context to obtain the corresponding classical Hamiltonian. Then the canonical momenta are
\be
p_1 = \frac{\der L_2}{\der \dot q_1} = a~{\dot q}_1 + \epsilon ~{\dot q}_2 ~~~~~ 
p_2 = \frac{\der L_2}{\der \dot q_2} = b~{\dot q}_2 + \epsilon ~{\dot q}_1 ~,
\ee
so that to first order in the small parameter $\epsilon$,
\be
{\dot q}_1 \approx \frac{p_1}{a} - \epsilon ~ \frac{p_2}{ab} ~~~~~~~~~~~~~~~~ 
{\dot q}_2 \approx \frac{p_2}{b} - \epsilon ~ \frac{p_1}{ab} ~,
\ee  
which, inserted in the Legendre transform of $L_2$, delivers the coupled Hamiltonian
\be
H_2(p_1, p_2, q_1, q_2) = \frac{p_1^2}{2a} +  \frac{p_2^2}{2b} - \epsilon ~ \frac{p_1p_2}{ab} + \omega_0^2 ~ (q_1^2 + q_2^2) ~.
\label{H2}
\ee
It is curious that the isotopic variation couples the symmetric and the antisymmetric variables by their momenta, instead  of the more
familiar position coupling.

In the example of a CO$_2$ molecule with isotopes $^{16}$O and $^{18}$O \cite{refCO2}, we take $k=1840$N.m$^{-1}$, which gives
\[
 \omega_0 \simeq 1.55\times10^{14} \textrm{rad.s}^{-1} ~~~ \mu_+ = \frac{9}{23} ~~~ \mu_-=\frac{8}{23} ~~~ a \simeq 0.1928 ~~~ b \simeq 0.7372 ~~~ \epsilon \simeq 0.0113.
\]
Then, taking the symplectic change of variable
\[
 p_1 \rightarrow \sqrt{\frac{\sqrt{2}}{\sqrt{a}\;\omega_0}} \;p_1 ~~~ p_2 \rightarrow \sqrt{\frac{\sqrt{2}}{\sqrt{b}\;\omega_0}} \;p_2 ~~~ q_1 \rightarrow \sqrt{\frac{\sqrt{a}\;\omega_0}{\sqrt{2}}} \; q_1 ~~~ q_2 \rightarrow \sqrt{\frac{\sqrt{b}\;\omega_0}{\sqrt{2}}} \;q_2, 
\]
one obtains (\ref{H2app}) with
\begin{eqnarray}
\omega_1 = \sqrt{\frac{2}{a}}\;\omega_0 \simeq 4.99\times 10^{14}\textrm{rad.s}^{-1} \nonumber \\
\omega_2 = \sqrt{\frac{2}{b}}\;\omega_0 \simeq 2.55\times 10^{14}\textrm{rad.s}^{-1} \nonumber \\
c = \frac{2\sqrt{2}}{(ab)^{3/4}}\;\omega_0 \;\epsilon \simeq 0.21\times 10^{14}\textrm{rad.s}^{-1}\nonumber .
\end{eqnarray}
Noticing that $c/\omega_1< c/\omega_2 \simeq 1/10$, it is legitimate to treat the coupling term in $c$ as a perturbation.

\section{Spectrum of the classical evolution matrix for the network of oscillators}
\label{eigen}
It is convenient to decompose the evolution matrix $\G_\Gamma$ of Sec. \ref{network} into a sum of two tensor products, that is, writing 
\be
\G_\Gamma  = \left( \matr 1_N+\frac{\alpha}{2\omega} \matr N\right) \otimes {\matr J_1} {\matr H_1} + \frac{\gamma}{2}\matr \Delta\otimes \matr 1_2
\ee
with $N\times N$ matrices $\matr N$ and $\matr \Delta$, and $2\times 2$ matrix $\matr H_1$, defined by
\be
\hspace{-2cm}
\matr N_{j,k} = \delta_{j,k+1}+\delta_{j,k-1}\;,
~~~~
\matr \Delta = \mD_{1,0,\ldots,0}\;,
~~~~
{\matr J_1} {\matr H_1} = \left(\begin{array}{cc} 0 & -m\omega^2 \cr
\frac{1}{m} & 0 \end{array} \right),
\ee
where $\delta_{j,k}$ is the Kronecker symbol.
Although ${\matr J_1} {\matr H_1}$ obviously commutes with $2\times 2$ identity $\matr 1_2$, $\matr 1_N+\frac{\alpha}{2\omega} \matr N$ does not commute with $\matr \Delta$. Still, the eigenvectors $\vct W$ of $\G_\Gamma$ can be factorized by the two eigenvectors $\vct \epsilon_{\pm}$ of a single harmonic oscillator,
\be
\vct W_\pm = \vct V_{\pm}\otimes \vct \epsilon_{\pm} ~~~ \textrm{with} ~~~ 
{\matr J_1} {\matr H_1}\vct \epsilon_{\pm} = \pm i\omega \vct \epsilon_{\pm}.
\label{defW}
\ee
Then, the eigenvalue problem of $\G_\Gamma$ can be reduced to two subproblems,
\begin{eqnarray}
\G_\Gamma \vct W_+ &=& \left( \matr 1+\frac{\alpha}{2\omega} \matr N\right) \vct V_+ \otimes i\omega \vct \epsilon_{+} + \frac{\gamma}{2}\matr \Delta\otimes \vct \epsilon_{+} \cr 
~ &=& \left[ i\omega\left( \matr 1+\frac{\alpha}{2\omega} \matr N\right) + \frac{\gamma}{2}\matr \Delta \right] \vct V_+ \otimes \vct \epsilon_{+} \cr
~ &=& \lambda_{+} \vct W_+ \cr
\G_\Gamma \vct W_- &=& \left( \matr 1+\frac{\alpha}{2\omega} \matr N\right) \vct V_- \otimes (-i\omega) \vct \epsilon_{-} + \frac{\gamma}{2}\matr \Delta\otimes \vct \epsilon_{-} \cr 
~ &=& \left[ -i\omega\left( \matr 1+\frac{\alpha}{2\omega} \matr N\right) + \frac{\gamma}{2}\matr \Delta \right] \vct V_- \otimes \vct \epsilon_{-} \cr
~ &=& \lambda_{-} \vct W_- .
\label{eigenW}
\end{eqnarray}
Noticing that $\matr 1$ commutes with anything, we can conclude that 
\begin{eqnarray}
\vct V_+ \textrm{~~~is~eigenvector~of~~~} \matr N - i\frac{\gamma}{\alpha} \matr \Delta\cr
\vct V_- \textrm{~~~is~eigenvector~of~~~} \matr N + i\frac{\gamma}{\alpha} \matr \Delta.
\label{eigenV}
\end{eqnarray}
Defining $\sigma =\pm i\frac{\gamma}{\alpha}$, both problems can be formulated as finding the spectrum of the $N\times N$ matrix $\matr T=\matr N + \sigma \matr \Delta$, which has the form
\be
\matr T_{j,k} = \delta_{j,k+1} + \delta_{j,k-1} + \sigma \delta_{j,1}\delta_{k,1}\; .
\ee
$\matr T$ is ``almost'' a T\oe plitz matrix, that is, it is a T\oe plitz matrix for $\sigma=0$. 
Its eigenvector, for eigenvalue $\nu$, is a finite sequence $\left\{a_n\right\}_{1,\ldots,N}$ having a linear recurrence relation of order $2$, whose solution is $a_n = K_1\;r^n+K_2\;r^{-n}$, with $r^2-\nu r+1=0$. Hence the eigenvalue $\nu$ is of the form $\nu = r+\frac{1}{r}$, with $r$ solution of $a_{N+1}=K_1\;r^{N+1}+K_2\;r^{-(N+1)}=0$, which, after setting $K_1$ and $K_2$ according to initial conditions $a_1=1$ and $\sigma + a_2=\nu$, boils down to
\be
r^{2N+2} - \sigma r^{2N+1} +\sigma r - 1=0.
\label{eqr}
\ee
For $\sigma=0$, the solution is $r=e^{i\frac{k\pi}{N+1}}$, with $k=1,2,\ldots,N$, thus recovering the T\oe plitz spectrum $\nu = 2\cos{\frac{k\pi}{N+1}}$. We skip values $k=N+2,\ldots,2N-1$ which give the same $\nu$, and $k=0$ or $k=N+1$ which give $r=\pm 1$, implying $\sigma=\pm 1$, which is not possible. In the small $|\sigma|$ limit, corresponding to an internal coupling which is much stronger than the coupling to the environment, we expect $r$ to be a $\sigma$ expansion around its T\oe plitz value,
\be
r = e^{i\frac{k\pi}{N+1}} + \sigma r^{(1)} + \sigma^2 r^{(2)} + \ldots,
\label{expr}
\ee
which gives, at first order in $\sigma$, after introducing (\ref{expr}) in (\ref{eqr}) and the relation $\nu=r+1/r$,
\be
\nu_k = 2\cos{\left(\frac{\pi k}{N+1}\right)} + 2\sigma \frac{\sin^2{\left(\frac{k\pi}{N+1}\right)}}{N+1} + \GO(\sigma^2).
\ee
But $\nu$ is only the eigenvalue of $\matr N + \sigma \matr \Delta$, whereas we want to solve (\ref{eigenW}), which is twofold: (i) $\lambda_{+}$ is an eigenvalue of $i\omega\matr 1+i\frac{\alpha}{2}\left( \matr N + \sigma \matr \Delta \right) $ with $\sigma = -i\frac{\gamma}{\alpha}$; (ii) $\lambda_{-}$ is an eigenvalue of $-i\omega\matr 1-i\frac{\alpha}{2}\left( \matr N + \sigma \matr \Delta \right)$ with $\sigma = i\frac{\gamma}{\alpha}$.

Hence, the eigenvalues $\lambda_{k\pm}$ of $\G_\Gamma$ are finally, at first order in $\frac{\gamma}{\alpha}$,
\be
 \lambda_{k\pm} = \pm i\left( \omega + \alpha \cos{\left(\frac{\pi k}{N+1}\right)} \right) + \frac{\gamma}{2(N+1)} \sin^2{\left(\frac{k\pi}{N+1}\right)} + \GO\left(\frac{\gamma^2}{\alpha}\right).
\ee
On the other hand, the normalized unperturbed eigenvectors are
\be
 \vct V_{k}^{(0)} = \sqrt{\frac{2}{N+1}}\left(\sin{\left(\frac{k\pi}{N+1}\right)},\sin{\left(\frac{2k\pi}{N+1}\right)},\ldots,\sin{\left(\frac{nk\pi}{N+1}\right)},\ldots,\sin{\left(\frac{Nk\pi}{N+1}\right)}\right).
\label{Vk0}
\ee

\section*{Acknowledgments}
We thank R. Vallejos for stimulating discussions.
Partial financial support from the 
National Institute for Science and Technology--Quantum Information
and CNPq (Brazilian agencies) is gratefully acknowledged.

\end{document}